\documentclass[prc,twocolumn,showpacs,superscriptaddress]{revtex4}

\usepackage{graphicx}%
\usepackage{dcolumn}
\usepackage{amsmath}
\renewcommand\vec[1]{\boldsymbol{#1}}

\def\Journal#1#2#3#4{{#1} {\bf #2}, #3 (#4)}

\def\JPA{{J.\ Phys.}\ A}

\def\NPB{{Nucl.\ Phys.}\ B}
\def\PLB{{Phys.\ Lett.}\  B}
\def\PRL{Phys. Rev. Lett.}

\def\PRD{{Phys.\ Rev.}\ D}
\def\PR{{Phys.\ Rev.}}
\def\PRT{{Phys.\ Rep.}}

\def\kvec{{\vec k}}
\def\absum{\langle\alpha\beta\rangle}
\def\thetakk{\vartheta_{\vec k \vec k'}}

\begin{document}

\preprint{}

\title{Colour superconductivity in finite systems}

\author{Paolo Amore}
\thanks{Present address: Facultad de Ciencias, Universidad de Colima, Mexico}%
\email{paolo@ucol.mx}
\affiliation{Theoretical Physics Group, Department of Physics and Astronomy, University of Manchester, 
Manchester M13 9PL, U.K.}
\affiliation{Department of Physics, UMIST, P.O. Box 88, Manchester M60 1QD, U.K.}
\author{Michael C. Birse} 
\affiliation{Theoretical Physics Group, Department of Physics and Astronomy, University of Manchester, 
Manchester M13 9PL, U.K.}
\author{Judith A. McGovern}
\affiliation{Theoretical Physics Group, Department of Physics and Astronomy, University of Manchester, 
Manchester M13 9PL, U.K.}
\author{Niels R. Walet} 
\affiliation{Department of Physics, UMIST, P.O. Box 88, Manchester M60 1QD, U.K.}

\date{\today}

\begin{abstract}
In this paper we study the effect of finite size on the two-flavour 
colour superconducting state. As well as restricting the quarks to a
box, we project onto states of good baryon number and onto colour singlets,
these being necessary restrictions on any observable ``quark nuggets''.
We find that whereas finite size alone has a significant effect for very small
boxes, with the superconducting state often being destroyed, the effect of projection 
is to restore it again.  The infinite-volume limit is a good approximation 
even for quite small systems. 
\end{abstract}

\maketitle


\section{Introduction}

It is widely believed that  quarks, the basic building blocks
of strongly interacting matter, lose their confinement at some large
density or temperature, at or near  the chiral phase
transition. Many models, for example the instanton liquid model,
suggest that the unconfined quarks exert an attractive force on each other.
The BCS mechanism predicts that an arbitrarily weak attractive force leads 
to an instability of the Fermi-surface, giving rise to
a colour superconductor \cite{BL84}. This new phase of quark matter
 has been the subject of  many papers in recent years;
see Ref.~\cite{RW00} for a detailed list of references.

The interplay of the order parameter with the flavour symmetry
(either SU(2) or SU(3), since only the lightest
quarks are relevant) can generate many  interesting
phases.  These include the two-flavour
colour superconductor (2SC)  which pairs quarks of different flavour and 
colour and leaves one colour unpaired \cite{ARW98};
the colour flavour locked phase (CFL) which is possible in the three-flavour case \cite{ARW99};
and the crystalline colour superconducting phase \cite{ABR01}. 
In general, a robust colour superconducting gap of the order of $100 \ {\rm MeV}$
has been predicted by  calculations
which use different interactions and different superconducting states. 
The density at which the maximum pairing occurs is several times nuclear density.

An important question is what happens in finite systems. If we ever
succeed in creating such a phase in the lab, it  must 
occur in finite, colour-singlet, states. Equally, if we have a
nugget of quark matter at the centre of a neutron star, we expect it to
be colour neutral, since it originates from the compression of a
colour neutral object and is surrounded by colour neutral matter.

The techniques with which to discuss such phenomena have long been used in nuclear physics,
where the superfluid nature of finite nuclei should not be described
by the number-non-conserving BCS state, but by a projected one \cite{RS80}.
The additional complication here is the requirement of colour neutrality, since
the Cooper pairs in 2SC superconductivity are colour anti-triplets.  Hence an
additional projection onto a colour-singlet state is required.  In contrast
the Cooper pairs in nuclear physics already carry zero angular momentum.
 
In this work we will use the model of Ref.~\cite{ARW98} and work at zero temperature.  
We limit our study to the simplest state, the 2SC. We model the effects of finite
size by confining the quarks to a cubic box with antiperiodic boundary conditions.
As a result boundary effects are not included.
After projection the exact equations for the superconducting gap are very involved, so in 
this work we confine ourselves 
to a one-parameter variation.  With these limitations, we find that the superconducting
gap persists after, though not always before, projection in systems of finite size.

Finite size effects in superconducting matter have also been considered in ref.~\cite{Madsen},
where it was found that the stability of small lumps of CFL matter was enhanced.  In that
paper however the emphasis was on thermodynamic effects; shell structure and projection
onto states of good quantum numbers were not considered.

The paper is organised as follows: in Section \ref{sec:model} we describe the model 
and fix the notation; in Section \ref{sec:numb_proj} we 
apply the  projection over baryon number to the 2SC and obtain an expression 
for the expectation value of the energy in this projected state; in Section \ref{sec:colour_proj}
we apply the projection over the colour singlet to the 2SC and generalise 
the expression given in the previous section. In  Section \ref{sec:res} we give the 
numerical results, and finally, in Section \ref{sec:concl} we state our conclusions. 
The reader will find technical details and some useful formulae
in the Appendices \ref{app:1} and \ref{app:2}.

\section{The model}\label{sec:model}

We wish to investigate the effects of 
finite size, definite baryon number and colour neutrality on the colour superconducting
state. As mentioned in the introduction, a variety of models have been
so far used to study the emergence of colour superconductivity. We follow
ref.~\cite{ARW98} and use a standard NJL model with massless quarks
carrying two flavours and three colours. We leave for future work the study of
more complex scenarios.

Following ref.~\cite{ARW98} we model the attractive interaction
responsible for the pairing of quarks at large densities with a
contact four-quark interaction whose form is suggested by the instanton-induced
interaction. The Hamiltonian then has the form 
$\hat H=\hat H_0+\hat{H}_{\text{int}}$ with
\begin{widetext}
\begin{eqnarray}
\hat{H}_0 &=&-\int \! d^3x \left(\overline{\psi}_{L i \alpha}(x) \,i \vec\alpha\cdot\vec\nabla \psi_{L i \alpha}(x) 
+\overline{\psi}_{R i \alpha}(x) \,i  \vec\alpha\cdot\vec\nabla \psi_{R i \alpha}(x)\right),\nonumber\\
\hat{H}_{\text{int}} &=& - K \ \Xi_{k l ; \alpha \beta \gamma \delta} \int \! d^3x \,
\overline{\psi}_{R 1 \alpha}(x)  \psi_{L k \gamma}(x) 
\overline{\psi}_{R 2 \beta}(x)  \psi_{L l \delta}(x)  + \text{h.c.},
\label{eq:hint1}
\end{eqnarray}\end{widetext}
where flavour and colour indices are Latin and Greek respectively, and 
the tensor $\Xi_{k l ; \alpha \beta \gamma \delta}$ is given by
\begin{eqnarray}
\Xi_{k l ; \alpha \beta \gamma \delta} \equiv \epsilon_{k l} \,
\left( 3 \delta_{\alpha \gamma} \delta_{\beta \delta} - 
\delta_{\alpha \delta} \delta_{\beta \gamma} \right) \ . 
\end{eqnarray}

At zero density the quarks acquire a dynamical mass, through the generation of a chiral 
condensate. This mass drops as the density increases, reaching  zero at the chiral phase transition.
At higher densities the colour condensate which signals the onset of colour superconductivity
forms.

In order to regularise the contributions stemming from large momenta, 
the authors of  Ref.~\cite{ARW98} have supplemented the interaction (\ref{eq:hint1}) 
with a form factor, to be associated with each quark of a given momentum.
We choose to use a form factor of the form
\begin{eqnarray}
F(k) &=& \frac{1+e^{-\varepsilon}}{1+e^{\varepsilon (k^2-\Lambda^2)/\Lambda^2}} \ ,
\label{eq:formfactor}
\end{eqnarray}

The parameter $\varepsilon$ determines the sharpness of the cut-off, and has been
taken to be 10.  This leaves two free parameters, $\Lambda$ and 
the coupling constant $K$ in the interaction.  These are fixed at zero density by requiring the  
pion decay constant to have its physical value, and by choosing the dynamically generated quark mass 
to be 400~MeV; this gives $\Lambda = 700$~MeV and $K=1.755\times 10^{-05}$~MeV$^{-2}$.

The modes of the fields are quantised in the standard way, and we
define the creation operators  
$\hat a_{(L,R) i \alpha}^\dagger(\kvec)$, 
$\hat b_{(L,R) i \alpha}^\dagger(\kvec)$, 
$\hat c_{(L,R) i \alpha}^\dagger(\kvec)$ 
for
particles, antiparticles and holes relative to the Fermi sea, $|k_F\rangle$,
which can thus be written as 
\begin{equation}
| k_F \rangle =\prod_{\alpha i \kvec}\theta(k_F-k) \hat{a}^\dagger_{Li\alpha}(\kvec)\hat{a}^\dagger_{Ri\alpha}(\kvec)
|0\rangle\ .
\end{equation}
We choose to work with states in which the Fermi momenta $k_F$ for all three colours are equal.
In fact the properties of the BCS state are independent of the initial Fermi momentum for the 
condensing quarks.  However colour projection becomes involved if $k_F$ is different for the
spectator quarks.

The Fermi sea is the natural choice of ground state at low to intermediate density.
At large densities the interaction (\ref{eq:hint1}) is responsible for the condensation
of Cooper pairs of quarks, 
and we get a new form for the wave function,
\begin{eqnarray}
| \Psi \rangle &=& \hat{G}_L \ \hat{G}_R \ | k_F \rangle,
\label{eq:wf1}
\end{eqnarray}
where
\begin{align}
\hat{G}_L& =\nonumber\\
 \prod_{\absum\kvec}\! &\left[ \cos\theta_A^L(k) + 
\epsilon_{\alpha \beta 3}  \sin\theta_A^L(k) e^{i \xi_A^L(k)} 
\hat{a}_{L 1\alpha}^\dagger(\kvec) \hat{a}_{L 2\beta}^\dagger(-\kvec) \right] \nonumber \\ 
 \prod_{\absum\kvec}\! &\left[ \cos\theta_B^R(k) + 
\epsilon_{\alpha \beta 3} \sin\theta_B^R(k) 
e^{i \xi_B^R(k)} \hat{b}_{R 1\alpha}^\dagger(\kvec)
\hat{b}_{R2\beta}^\dagger(-\kvec) \right] \nonumber \\
\prod_{\absum\kvec}\! &\left[ \cos\theta_C^R(k) + 
\epsilon_{\alpha \beta 3} \sin\theta_C^R(k) 
e^{i \xi_C^R(k)} \hat{c}_{R 1\alpha}^\dagger(\kvec)
\hat{c}_{R2\beta}^\dagger(-\kvec) \right] \nonumber \\
&\equiv\hat{G}_{L A} \,\hat{G}_{L B} \,\hat{G}_{L C}
\label{eq:GL}
\end{align}
and similarly for $\hat{G}_R$,  with the substitutions $L \leftrightarrow R$.
The sum denoted $\absum$ is taken over ${\alpha,\beta}={1,2}$ and ${2,1}$ only;
the BCS angles $\theta_{(A,B,C)}^{(L,R)}(k)$ and $\xi_{(A,B,C)}^{(L,R)}(k)$ are 
to be determined variationally. 

The BCS wave function $|\Psi\rangle$ of Eq.~(\ref{eq:wf1}) describes the condensation of pairs 
of quarks with the same helicity, but antisymmetric in flavour and in
colour, as can be seen by inspection of Eq.~(\ref{eq:GL}). 
The condensate carries a colour-$\bar 3$  order parameter.
Only two colours, by convention  $1$ and $2$ (or red and green), condense, 
leaving the third colour inactive. This corresponds to a spontaneous symmetry breaking.  

The BCS angles are obtained by minimizing the thermodynamic potential 
\begin{equation}
F=\langle \Psi |\hat{H}-\mu \hat{N}|\Psi\rangle.
\label{eq:Gibbs}
\end{equation}
The phases $\xi$
are fixed by requiring that the expectation value of $H_{\rm int}$ is maximally
attractive; this gives
\begin{eqnarray}
   \xi^L_C =    -\xi^L_A =  -\xi^L_B &=& \pi/2 \nonumber\\
    \xi^L_{(A,B,C)} + \xi^R_{(A,B,C)} &=& \pi.
\end{eqnarray}
Minimising with respect to the functions $\theta$  one finds the gap $\Delta$ 
and the angles $\theta_{(A,B,C)}(k)$ for particles, antiparticles and holes 
respectively to be given by the solutions of the coupled equations
\begin{align}
\tan &\,2\theta_{(A,B,C)}(k) \nonumber\\
&= F^2(k) \ \Delta \ \left(\frac{\theta(k-k_F)}{k-\mu} , 
\frac{1}{k+\mu} , \frac{\theta(k_F-k)}{\mu-k} \right) \ ,
\label{eq:solu}
\end{align}
and
\begin{align}
\Delta = \frac{K}{\pi^2} \int_0^\infty\! & k^2dk \,F^2(k)
  \biggl[
\theta(k-k_F) \, \sin 2\theta_A(k) +\nonumber\\&\sin 2\theta_B(k) +
\theta(k_F-k) \, \sin 2\theta_C(k) \biggr] \ .
\label{eq:gap}
\end{align}

\begin{figure}
\centerline{\includegraphics[width=7cm]{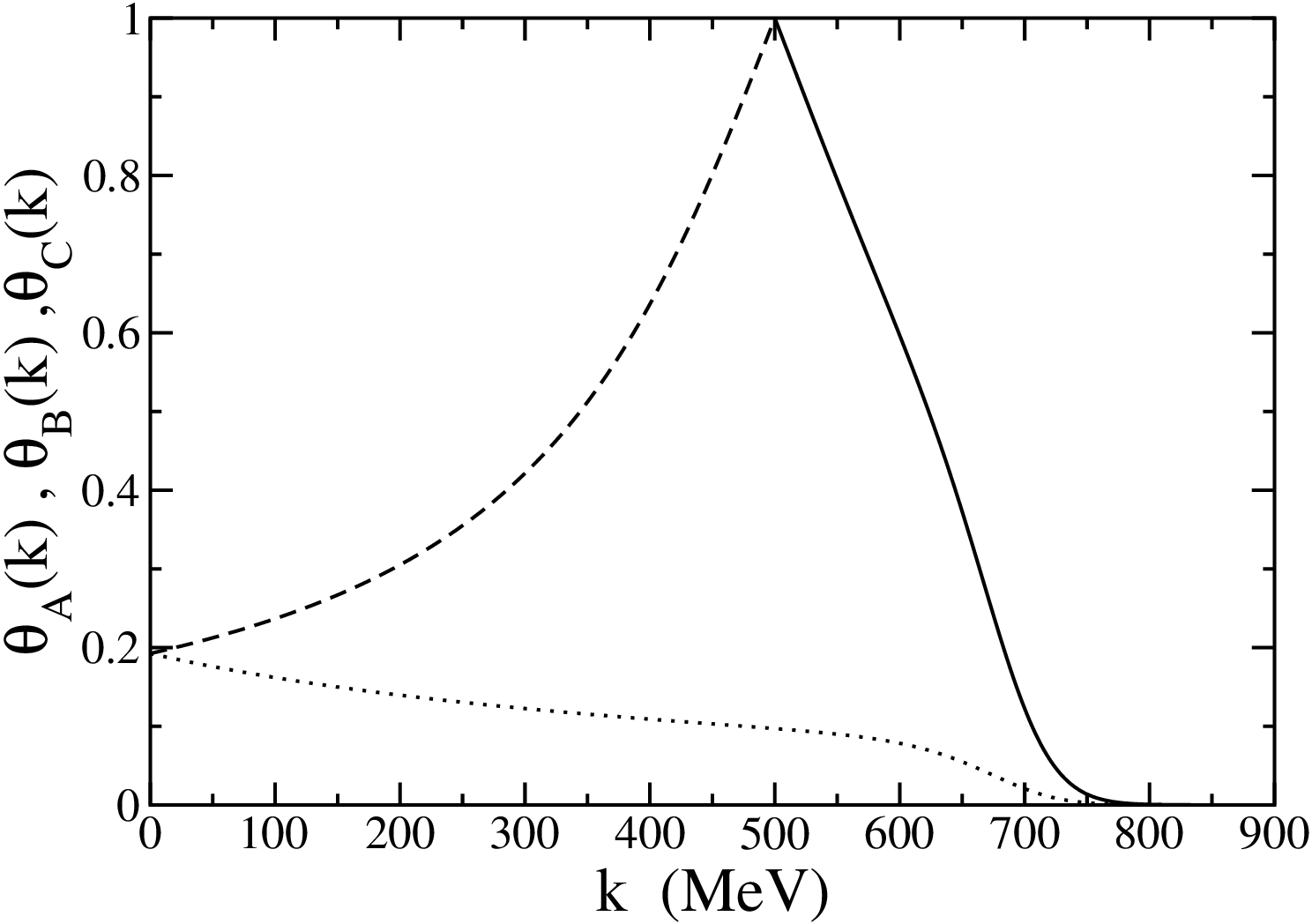}}
\caption{Variational solution for the infinite system at $\mu = 500 \ {\rm MeV}$. The angles are in
units of $\pi/4$. The solid, dotted and dashed lines correspond to particle, antiparticle
and hole contributions respectively.}
\label{fig:paperfig2}
\end{figure}

In Fig.~\ref{fig:paperfig2} we display the solutions for the BCS angles $\theta_A$, $\theta_B$ and
$\theta_C$ in the infinite volume for a density corresponding to a chemical potential 
$\mu = k_F=500 \ \rm MeV$. The solid, dotted and dashed lines correspond to the particle, 
antiparticle and hole contributions respectively. The solutions
quickly fall off at momenta larger than the cutoff $\Lambda$ as a result of the action of the 
form factor. Pairing is most effective at the Fermi surface, $k \approx \mu$.

The BCS wave function (\ref{eq:wf1}) does not have a definite baryon number 
(the condensate is a superposition of states with different number of pairs) 
and it is not in a colour singlet state. 
However, the number of particles is fixed on average 
by the Lagrange multiplier $\mu$ in Eq.~(\ref{eq:Gibbs}).
Because the fluctuations in the baryon number behave as $\sqrt{N}$, 
they become negligible as a fraction of the baryon number when the condensate 
contains a large number of particles.

\begin{figure}
\centerline{\includegraphics[width=7cm]{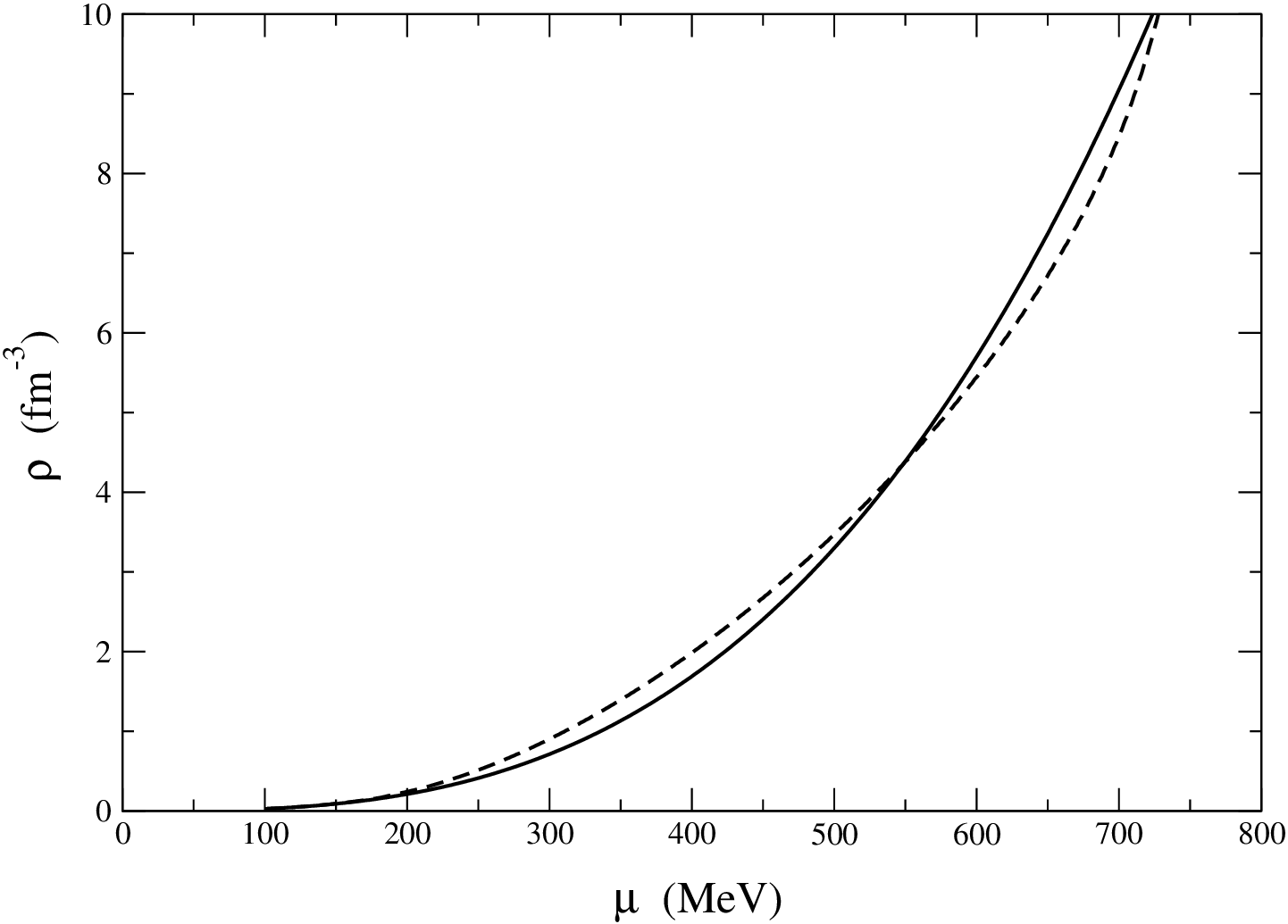}}
\caption{Quark density in the Fermi gas (solid) and in the BCS condensate (dashed) as a function
of the chemical potential.}
\label{fig:paperfig3}
\end{figure}

In Fig.~\ref{fig:paperfig3} we show the quark density as a function of the chemical potential
for a Fermi gas of quarks (solid line) and for the two-flavour BCS condensate (dashed line).
At small density the condensate contains more particle pairs than antiparticle or hole pairs; as
the density grows, the reverse is true, since the particle states start to be suppressed from the
form factor and more (Fermi) hole states become available.

In this paper we are interested in the effect of finite size on the colour 
superconductor.  To do this we consider the quarks to be in a cubic box with
antiperiodic boundary conditions. We start with a Fermi sea in which all levels with $k<\mu$
are occupied, and consider the projected and unprojected BCS states built on this.
For simplicity we only consider filled shells.

\begin{figure}
\centerline{\includegraphics[width=7cm]{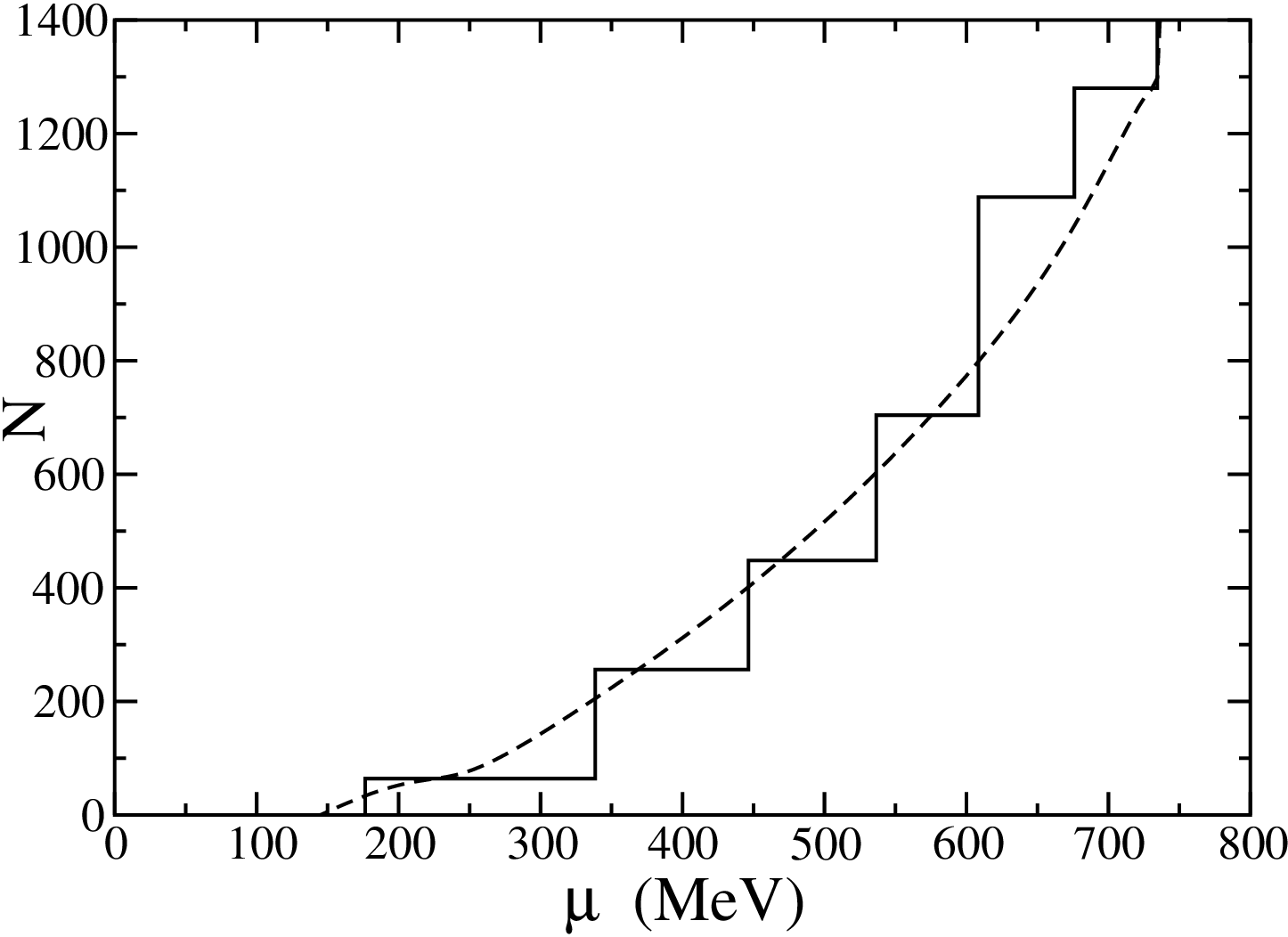}}
\caption{Solid line:   number of red and green particles in the Fermi sea as a function of 
the chemical potential
$\mu$ for a box of size $L = 6 \text{ fm}$. Dashed line: expectation value of the number of red and green 
particles in the BCS condensate  as a function of $\mu$.
At the intersections between the two curves  the  number of pairs of particles in the BCS state equals 
on average the number of pairs of antiparticles and holes. }
\label{fig:paperfig6}
\end{figure}

In Fig.~\ref{fig:paperfig6} we have plotted the quark density as a function of $\mu$
for the two active colours in the Fermi gas
(continuous line) and in the BCS condensate (dashed line), in a finite box of size $L = 6 \ \text{fm}$.
Notice that the dashed curve does not depend on $k_F$; in fact 
only quantities involving the third (inactive) colour will depend on $k_F$.  Because we work
with equal values of $k_F$ (that is the same filled shells) for all three colours,
at the intersection between the two curves the box will contain on average the same number 
of red, green and blue particles.  These are the points at which we will project out colour singlets.

In the following sections we study the occurrence of colour superconductivity
in systems of finite size and with a fixed number of particles.
We first project  the BCS wave 
function  onto states of definite baryon number (Section \ref{sec:numb_proj})
and then onto colour-singlet states (Section \ref{sec:colour_proj}).

\section{Projection onto definite baryon number}
\label{sec:numb_proj}

There are various ways to solve the problem of picking a
component of fixed particle number from a BCS state.
The most efficient ones are based on the calculation
of a contour integral; we follow here 
the method of residues of Ref.~\cite{DMP64}.
This will be used to project the colour 
superconducting  state considered in Eq.~(\ref{eq:wf1}) onto a state with 
a fixed number of particles.

We start by defining the rotated BCS state
\begin{eqnarray}
| \Psi(\zeta) \rangle&=&e^{i(\hat N -N)\phi/2}|\Psi\rangle \nonumber \\
&=&\zeta^{(\hat N -N)/2}|\Psi\rangle
\end{eqnarray}
where $\hat N$ is the number operator, $N$ is the number of particles in the Fermi
sea and $\zeta$ is a complex number on the unit circle with phase $\phi$.
The rotated state can be created directly from the Fermi sea by the rotated 
operators $\hat{G}_{(L,R)}(\zeta)\equiv \zeta^{ \hat N/2}\hat{G}_{(L,R)}\zeta^{- \hat N/2}$:
\begin{equation}
| \Psi(\zeta) \rangle=\hat{G}_L(\zeta) \hat{G}_R(\zeta) | k_F \rangle,
\end{equation}
where 
\begin{widetext}
\begin{eqnarray}
\hat{G}_L(\zeta) &=& \prod_{\absum\kvec} \left[ \cos\theta_A^L(k) + 
\zeta  \epsilon_{\alpha \beta 3}  
\sin\theta_A^L(k) e^{i \xi_A^L(k)}  \hat{a}_{L 1\alpha}^\dagger(\kvec)
 \hat{a}_{L 2\beta}^\dagger(-\kvec) \right] \nonumber \\
&\times& \prod_{\absum\kvec} \left[ \cos\theta_B^R(k) + 
\zeta^*  \epsilon_{\alpha \beta 3}  \sin\theta_B^R(k) 
e^{i \xi_B^R(k)}  \hat{b}_{R 1\alpha}^\dagger(\kvec)
 \hat{b}_{R2\beta}^\dagger(-\kvec) \right] \nonumber \\
&\times& \prod_{\absum\kvec}  \left[ \cos\theta_C^R(k) + 
\zeta^*  \epsilon_{\alpha \beta 3}  \sin\theta_C^R(k) 
e^{i \xi_C^R(k)}  \hat{c}_{R 1\alpha}^\dagger(\kvec)
 \hat{c}_{R2\beta}^\dagger(-\kvec) \right] \nonumber \\
&\equiv&  \prod_{\absum\kvec} \hat{G}_{L A \alpha \beta}(\zeta,\kvec) 
 \prod_{\absum\kvec} \hat{G}_{L B \alpha \beta}(\zeta,\kvec) 
 \prod_{\absum\kvec} \hat{G}_{L C \alpha \beta}(\zeta,\kvec),  
\label{eq:GLz}
\end{eqnarray}
\end{widetext}
and similarly for $\hat{G}_R(\zeta)$.
Because the creation of antiparticles or holes lowers the baryon number, 
the corresponding terms in  $\hat{G}$  contain $\zeta^*$ or,
equivalently, $1/\zeta$.

The number-projected wave function is produced from an appropriate
superposition of rotated states:
\begin{eqnarray}
| \Psi_{n} \rangle &=&  C_n \oint \frac{d \zeta}{\zeta^{n+1}}  
 | \Psi(\zeta) \rangle
\label{eq:wf2}
\end{eqnarray}
where $n$ is the number of pairs.  

The only operators whose expectation values in the number-projected state are non-vanishing
are those which are particle-number conserving and so commute with the number 
operator $\hat N$.  This enables us to write
\begin{equation}
\langle \Psi_{n}|\hat O | \Psi_{n} \rangle = \frac{\langle \Psi|\hat O | \Psi_{n} \rangle} 
{\langle \Psi | \Psi_{n} \rangle}
\label{eq:expect}
\end{equation}

A convenient way to calculate these matrix elements is to find the Thouless operator
$\hat S(\zeta)$ which maps  $| \Psi \rangle\equiv | \Psi(1) \rangle$ into  
$| \Psi(\zeta) \rangle$ \cite{RS80}: 
\begin{eqnarray}
\hat{S}(\zeta)  | \Psi \rangle =  | \Psi(\zeta) \rangle .
\label{eq:Thouless1}
\end{eqnarray}
The explicit expression for $\hat{S}$ is given in  Appendix \ref{app:1}. 
There we also give the definitions of the quasi-particle operators $\alpha(\zeta,\kvec), 
\beta(\zeta,\kvec),\gamma(\zeta,\kvec)$
which annihilate the state $|\psi(\zeta) \rangle$. 

The Thouless operator can be expanded in a series of terms containing zero, two, four, \ldots\ 
quasiparticle creation operators:
\begin{eqnarray}
\hat{S}(\zeta) = \sum_{n=0}^{\infty}  \hat{S}^{(n)}(\zeta).
\label{eq:expand}
\end{eqnarray}
Depending on the form of the interaction,
only a limited number of these terms will contribute to
the matrix elements of the Hamiltonian.
Luckily, the interaction (\ref{eq:hint1}) selects only the first two terms in the expansion. 
For left-handed particles these are
\begin{eqnarray}
S_{L A}^{(0)}(\zeta)&=& \prod_{\absum \kvec} 
\left( \cos^2\!\theta_A^L(k) + \zeta \sin^2 \!\theta_A^L(k) \right), \nonumber\\
\hat{S}_{L A}^{(1)}(\zeta)&=&  S_{L A}^{(0)}(\zeta)
\nonumber \\
&&\times \sum_{\gamma \delta k} \frac{(\zeta-1) \ \epsilon_{\gamma \delta 3}  \ \sin\theta_A^L(k) 
\cos\theta_A^L(k) e^{i \xi_A^L(k)}}{ \cos^2\!\theta_A^L(k) + \zeta \sin^2 \!\theta_A^L(k)}\nonumber\\&&
\qquad\times
  \hat{\alpha}_{L 1 \gamma}^\dagger(1,\kvec) \hat{\alpha}_{L 2 \delta}^\dagger(1,\kvec)  \ .
\end{eqnarray}

The function $S^{(0)}(\zeta)= \langle \Psi| \Psi(\zeta) \rangle$ can be expanded as a 
Taylor-Laurent series about $\zeta=0$, with real coefficients $d_n$.  These are related 
to the normalisation constant $C_n$ in Eq.~(\ref{eq:wf2}), by $2\pi i C_n=1/ \sqrt d_n$, and
to the expansion of the BCS state  $| \Psi \rangle$ in terms of states of definite particle number:
\begin{eqnarray}
| \Psi \rangle=\sum_n\sqrt{d_n}| \Psi_n \rangle.
\label{eq:comp}
\end{eqnarray}
Since $| \Psi \rangle$ has unit norm, the sum of the $d_n$ is unity.

From Eq.~(\ref{eq:expect}) we obtain the expectation value of the Hamiltonian in the number projected state: 
\begin{equation}
\langle  \Psi_n | \hat{H} | \Psi_n\rangle = - 2 \pi i \ |C_n|^2 \ \oint \frac{d\zeta}{\zeta^{n+1}} 
\langle \Psi | \hat{H} |  \Psi(\zeta) \rangle \ .
\end{equation}

Explicitly, we obtain for the non-interacting part of the Hamiltonian:
\begin{align}
\langle \Psi_n | \hat{H}_0 | \Psi_n \rangle & =  
\frac{4}{d_n}\!
\sum_{\vec k}  k  \nonumber \\
&\left\{ \theta(k_F-k)  
	\left[d_n+2 \cos^2\!\theta_C(k) \,	{\cal I}_{c,n}(\theta_C(k))\right]
\right.\nonumber\\
&+2 \,\theta(k-k_F) \sin^2 \!\theta_A(k)\, {\cal I}_{a,n}(\theta_A(k))  \nonumber \\
&+ \left. 2\sin^2 \!\theta_B(k) \,{\cal I}_{b,n}(\theta_B(k))
 \right\} \ .
\label{eq:h02}
\end{align}
In this expression the first term describes the hole contribution,
the second the particle one, and the last the antiparticle.
The first term ($d_n$) in the square brackets describes the spectator
colour; the second term reduces to twice this in the absence
of pairing (when $\theta_{(A,B,C)}=0$).
The interaction Hamiltonian leads to the  more complicated form
\begin{widetext}
\begin{eqnarray}
\lefteqn{\langle \Psi_n | \hat{H}_{\text{int}} | \Psi_n \rangle}\nonumber\\
&=&  - \frac{8  K}{\Omega  d_n} \sum_{\vec k} \sum_{\vec k'}  F^2(k) \
F^2(k') \nonumber \\
&&\times \left\{  \theta(k_F-k) \theta(k_F-k')  \sin^2 \!\frac{\thetakk}{2}   
\sin 2\theta_C(k)  \sin 2\theta_C(k')   \right. 
\left. {\cal J}_{b,n}(\theta_C(k),\theta_C(k'))
\right. \nonumber \\
&&+ \left. 2  \theta(k_F-k)  \cos^2\!\frac{\thetakk}{2}  \sin2 \theta_C(k) \
\sin 2\theta_B(k')   \right. 
\left. {\cal J}_{b,n}(\theta_C(k),\theta_B(k'))
\right. \nonumber \\
&&+ \left. 2  \theta(k_F-k) \theta(k'-k_F)  \cos^2\!\frac{\thetakk}{2} \
\sin 2\theta_C(k)  \sin 2\theta_A(k')  \right. 
\left. {\cal J}_{c,n}(\theta_C(k),\theta_A(k'))  \right. 
\nonumber \\&&+ 
\left.   \sin^2 \!\frac{\thetakk}{2}  \sin 2\theta_B(k)  \sin 2\theta_B(k')  \right. 
\left. {\cal J}_{b,n}(\theta_B(k),\theta_B(k'))
\right. \nonumber \\
&&+ \left. 2  \theta(k'-k_F)  \sin^2 \!\frac{\thetakk}{2}  \sin 2\theta_B(k)  
\sin2\theta_A(k')   \right. 
\left. {\cal J}_{c,n}(\theta_B(k),\theta_A(k'))
\right. \nonumber \\
&&+ \left.  \theta(k-k_F) \theta(k'-k_F) \sin^2 \!\frac{\thetakk}{2} 
\sin 2\theta_A(k)\sin 2\theta_A(k') \right. 
\left.  {\cal J}_{a,n}(\theta_A(k),\theta_A(k'))
\right\} \,  .
\label{eq:hint2}
\end{eqnarray}
\end{widetext}
The volume of the box is $\Omega\equiv L^3$.
The contour integrals are included in the definitions of ${\cal
I}_{(a,b,c),n}$ and ${\cal J}_{(a,b,c),n}$ (see Appendix A); $\thetakk$ is
the angle between $\vec{k}$ and $\vec{k}'$. The form factors in the
expression (\ref{eq:hint2}) suppress the contributions
coming from large momenta.

The expectation value of the number operator is obtained from Eq.~(\ref{eq:h02})
by dropping the single-particle energy $k$ which appears immediately after the summation sign
and swapping the sign of the antiparticle ($B$) term.

These expressions (\ref{eq:h02}) and (\ref{eq:hint2}) reduce to the expectation values in the 
unprojected BCS state if the integrals ${\cal I}$ and ${\cal J}$ and the constant $d_n$ are 
set to one.    The infinite-volume gap equations (\ref{eq:solu}) and (\ref{eq:gap}) 
are easily obtained  by minimising $\langle \hat{H}-\mu \hat{N}\rangle$ with respect to the angles $\theta$,
if the sums are replaced by integrals.
However in the finite volume projected  case, the equations are much more complicated.  
For a start, in the cubic box there
is no reason to expect the angles to depend only on the length and not on the direction of $\kvec$.
This introduces a dependence on the specific geometry that one might be quite happy to ignore.  However
even then the dependence of the  ${\cal I}$'s and ${\cal J}$'s on the $\theta$'s renders the equations
very complicated.  We do not attempt to solve these.  Instead we assume that the form of the 
$\theta$s is given as before by Eq.~(\ref{eq:solu}), and minimise the expression for 
$\langle \hat{H}-\mu\hat{N}\rangle$ obtained from Eqs~(\ref{eq:h02}) 
and (\ref{eq:hint2})  with respect to the gap $\Delta$ only.

This process of restricting the BCS angles to have the same form as in the unprojected state
will presumably introduce fewest errors where the projected state has a high overlap with
the unprojected state.  Thus although in principle we can project a state of a given baryon number
from any BCS state, in practice we choose values 
of $\mu $ where the expectation value of the number of pairs in the unprojected BCS state has the
value desired in  the projected state (see  Fig.~\ref{fig:paperfig6}).

In the BCS state there are non-vanishing expectation values for diquark condensates
such as $\psi_{R 1 1}(x) \psi_{R 2 2}(x)$.  However as this operator changes
the quark number by two units, 
it does not commute with the number operator $\hat N$ and the expectation value in the number-projected
state will vanish.  None-the-less, the square of the diquark condensate operators do not vanish, and they act as order
parameters in this case.

\section{colour projection}
\label{sec:colour_proj}

\begin{figure*}[t]
\centerline{
\includegraphics[width=7cm]{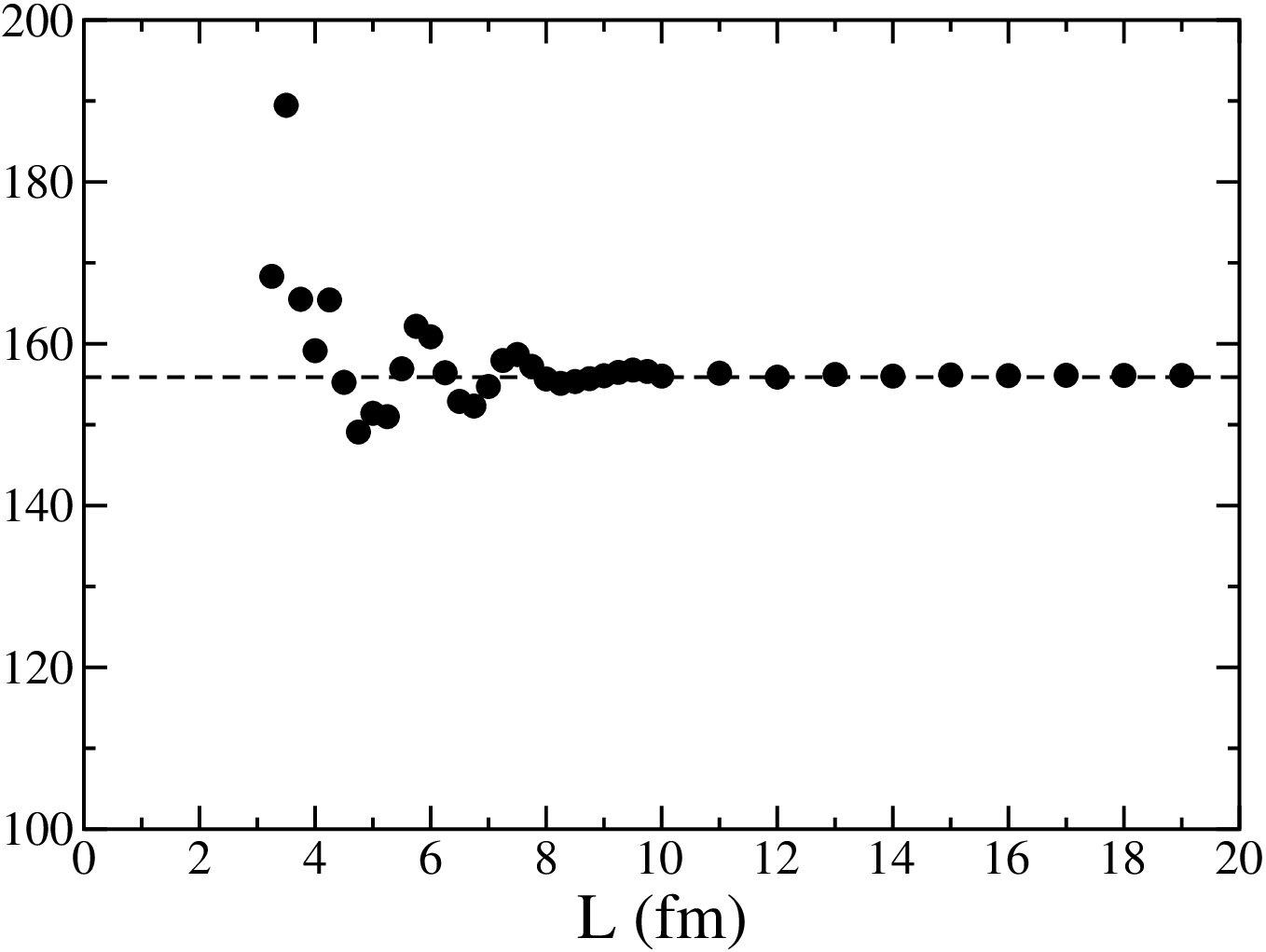}
\hspace{1cm}
\includegraphics[width=7cm]{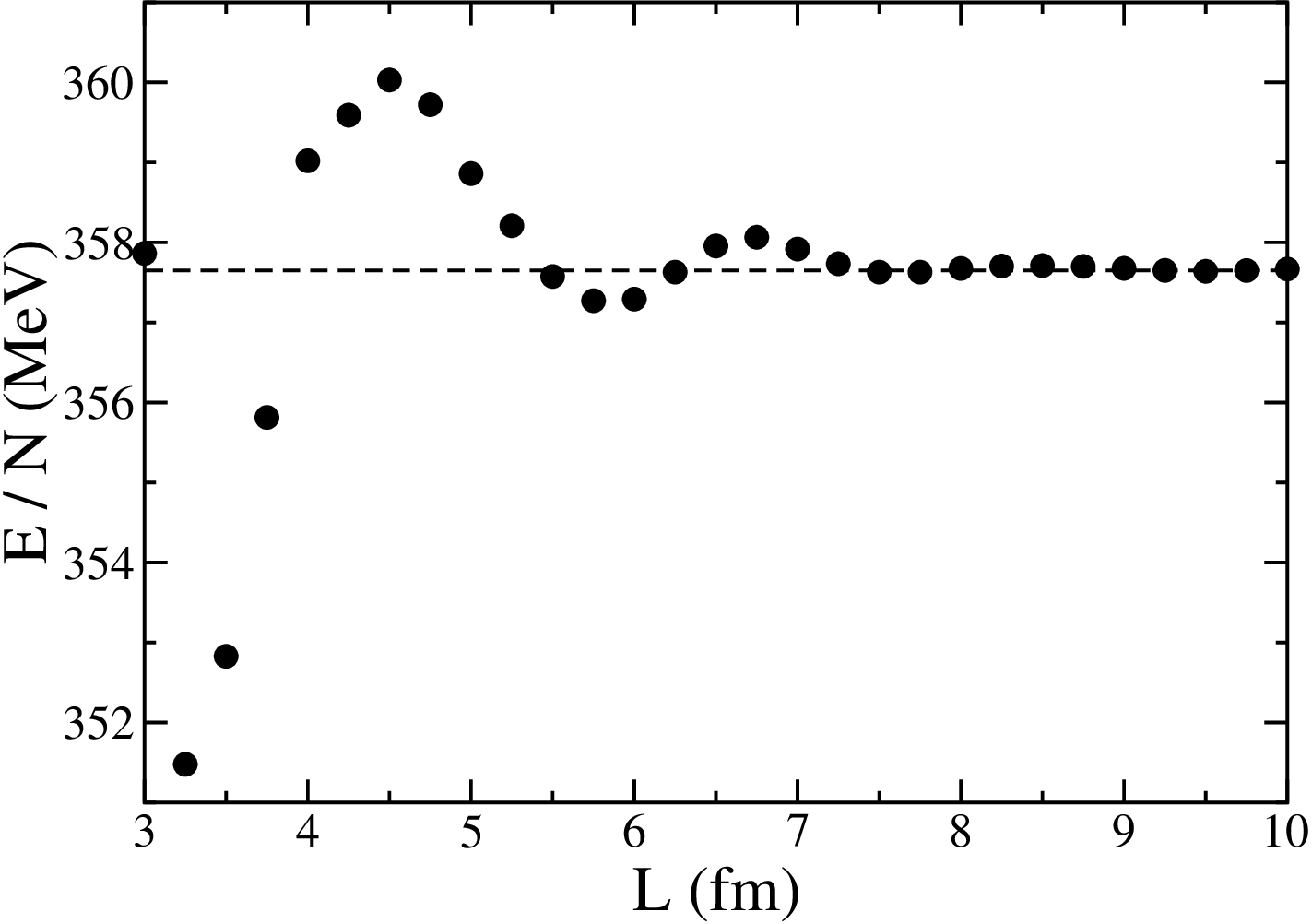}}
\caption{Approach of the gap (left) and energy per quark for the paired colours only (right)
to the infinite volume result (dashed line) 
at a fixed chemical potential $\mu = 500 \ {\rm MeV}$.}
\label{fig:fsize1}
\end{figure*}

We now turn to the issue of colour projection. As we have pointed out before, the colour-superconducting
state which forms at large densities must be a colour singlet. However, the state in Eq.~(\ref{eq:wf1})
does not fulfill this requirement, given the form of the operators $\hat{G}_{(L,R)}$.
For this reason we want to project the colour-singlet state from  Eq.~(\ref{eq:wf1}).
This requires us to integrate over
the group manifold:
\begin{eqnarray}
|\widetilde \Psi_n\rangle &=& \int \! d\Omega_g \,\hat{U}_g \,|\Psi_n\rangle \ .
\end{eqnarray}
Here $d\Omega_g$ is the volume element on the group manifold and 
$|\Psi_n\rangle$ is the number projected BCS state. 
$\hat{U}_g$ is a unitary operator which performs a rotation in the $SU(3)$ space.

Following \cite{Byrd98} we parametrize an element $g$ of SU(3) in the form
\begin{align}
g = e^{i \frac{\phi_a}{2} \lambda_3}  e^{i \frac{\phi_b}{2} \lambda_2}
  e^{i \frac{\phi_c}{2} \lambda_3} 
e^{i \frac{\phi}{2} \lambda_5}  e^{i \frac{\phi'_a}{2} \lambda_3}  e^{i \frac{\phi'_b}{2} \lambda_2}  
e^{i \frac{\phi'_c}{2} \lambda_3}  e^{i \frac{\phi_e}{2} \lambda_8} ,
\label{eq:rotation}
\end{align}
where $\lambda_i$ are the Gell-Mann matrices, and write the volume element as
\begin{align}
d\Omega_g = \frac{1}{2^8} &
\sin(\phi_b)  \sin(\phi'_b)  \sin(\phi)  \sin^2 \!\left(\frac{\phi}{2}\right) \nonumber\\ & \times d\phi_a \,
d\phi_b \, d\phi_c \, d\phi \, d\phi'_a \, d\phi'_b \, d\phi'_c \, d\phi_e \ .
\end{align}
To obtain the realization of these rotation operators when
acting on our quarks states, we replace the $\lambda$'s
with the transition operators 
$\hat{Q}_a = \int \! d^3x \, \psi^\dagger(x) \frac{\lambda_a}{2} \psi(x)$,
which are associated with the SU(3) group.

Because of the residual colour symmetry, the operators associated with the red-green SU(2)
subgroup annihilate the BCS state, and also commute with the number projection operation.
Thus
\begin{eqnarray}
\hat{Q}_{1,2,3} |\Psi_n \rangle &=& 0 .
\end{eqnarray}

We can simplify the algebra considerably by noting that
\begin{eqnarray}
\hat{Q}_8 &=& \int \! d^3x \, \psi^\dagger(x) \, \frac{\lambda_8}{2} \, \psi(x) 
\nonumber\\
&=&
\frac{1}{2} \left[ \hat{N}_{11} + \hat{N}_{22} - 2 \ \hat{N}_{33} \right] \ ,
\end{eqnarray}
where $1$, $2$ and $3$ are the values taken by the the colour
index. Since we are working with states built on Fermi seas which have equal
numbers of red, green and blue quarks, we see that $Q_8$ annihilates only 
the number-projected BCS state $|\Psi_0\rangle$ which contains zero net pairs
({\it i.e.}\ the number of pairs of holes and antiparticles is the same as the
number of pairs of particles).  We therefore restrict our projection to this
state.  

Thus the only operator which acts non-trivially  is $\hat{Q}_5$, which is
non-diagonal in the colour space and connects one of the two paired
colours with the remaining one. As a result
we are able to reduce the integration over the eight different
angles $\phi_i$ in Eq.~(\ref{eq:rotation}) down to an integration
over the single angle $\phi$ for the rotation generated by $\hat{Q}_5$.
The residual volume element is therefore
\begin{eqnarray}
d\Omega_5 &=& \frac{1}{2} \sin(\phi) \, \sin^2 \!\left(\frac{\phi}{2}\right)  d\phi \nonumber\\
&=& \sin^3\!\left(\frac{\phi}{2}\right) \, d\sin\!\left(\frac{\phi}{2}\right),
\label{eq:red}
\end{eqnarray}
where $0 \leq \phi \leq \pi$ is the angle associated with $\hat{Q}_5$.

As before, we want to project states with zero pairs from BCS states which have zero pairs
on average, so we are restricted to working only at certain values of $\mu$.  These are
the values at which the two curves of  Fig.~\ref{fig:paperfig6} cross.  For small boxes,
this is a significant restriction on the densities we can consider.

Just as in the case of number projection particle-number-conserving operators have non-zero 
expectation values, so when we project onto a colour singlet only colour-singlet operators
will have non-vanishing expectation values.  Such operators commute with the colour rotation, and so
we can write (cf.~ Eq.~(\ref{eq:expect}))
\begin{eqnarray}
\langle \widetilde\Psi_0 |\hat{O}|\widetilde\Psi_0 \rangle &=& 
\frac{\int \! d\Omega_g \, \langle \Psi_0 |\, \hat{O}\,\hat{U}_g \,|\Psi_0\rangle}
     {\int \! d\Omega_g \, \langle \Psi_0 |\,          \hat{U}_g\, |\Psi_0\rangle}  \nonumber\\
&=&  \frac{\int \! d\Omega_5 \, \langle \Psi_0 |\, \hat{O}\, e^{i \frac{\phi}{2} \hat{Q}_5}\,|\Psi_0\rangle}   
          {\int \! d\Omega_5 \, \langle \Psi_0 |\,           e^{i \frac{\phi}{2} \hat{Q}_5}\,|\Psi_0\rangle}\ .
\label{eq:expct_vl}
\end{eqnarray}

We can write  the colour-number projected state in terms of the colour rotated 
operators $\widetilde{{G}}_{(L,R)}(\zeta)$, 
\begin{eqnarray}
| \widetilde{\Psi}_0\rangle &=&  \tilde C_0 \oint \frac{d \zeta}{\zeta}  \int \!  d\Omega_5\,
 e^{i \frac{\phi}{2} \hat{Q}_5} \,  | \Psi(\zeta) \rangle \nonumber \\
&\equiv& \tilde C_0 \oint \frac{d \zeta}{\zeta} \int \! d\Omega_5\,
\widetilde{{G}}_L(\zeta,\phi)  \widetilde{{G}}_R(\zeta,\phi)  | k_F \rangle \ ,
\label{eq:wf4}
\end{eqnarray}
where $\widetilde{{G}}_{(L,R)}(\zeta,\phi) \equiv \hat{U}_g  \hat{G}_{(L,R)}(\zeta)  \hat{U}_g^\dagger$;
for example, the operators for the left-handed particles read 
\begin{widetext}
\begin{eqnarray}
\widetilde{{G}}_{L A 1 2}(\kvec,\zeta,\phi) &=& \cos\theta_A^L(k) +
\zeta \ \sin\theta_A^L(k) e^{i \xi_A^L(k)} \ 
\left( \cos\frac{\phi}{2} \  \hat{a}^\dagger_{L 1 1}(\kvec) - 
\sin\frac{\phi}{2} \ \hat{a}^\dagger_{L 1 3}(\kvec) \right) \hat{a}^\dagger_{L 2 2}(-\kvec) \ ,\nonumber \\
\widetilde{{G}}_{L A 2 1}(\kvec,\zeta,\phi) &=& \cos\theta_A^L(k) -
\zeta \ \sin\theta_A^L(k) e^{i \xi_A^L(k)} \ 
\hat{a}^\dagger_{L 1 2}(\kvec) \ \left( \cos\frac{\phi}{2} \  
\hat{a}^\dagger_{L 2 1}(-\kvec) - \sin\frac{\phi}{2} \ \hat{a}^\dagger_{L 2 3}(-\kvec) \right) \ ,
\label{eq:rot2}
\end{eqnarray}
\end{widetext}
(recall that the first 
numerical index on the creation operators is the flavour index, and the second is colour).
This clearly reduce to the original form Eq.~(\ref{eq:GLz}) for $\phi = 0$. 
By looking at the expressions (\ref{eq:rot2})  we see that, by acting with $\widetilde{{G}}_{L A 1 2}$ on the 
Fermi sea, one can create pairs of particles in the colours $12$ or in the colours $23$. 
The probability of each of these events will depend on the magnitude of the angle $\phi$; 
clearly, the two extrema of the interval (i.e. $\phi = 0$ and $\phi = \pi$) correspond to pure 
pairing in the $12$ and $23$ colours respectively.

Once again, calculations are most conveniently done with the aid of 
a Thouless operator, $\hat{W}$, which fulfills the relation
\begin{eqnarray}
 e^{i \frac{\phi}{2} \hat{Q}_5}|\psi(\zeta) \rangle &=&  
\hat{W}(\zeta, \phi) |\Psi\rangle. 
\end{eqnarray}
The expectation value (\ref{eq:expct_vl})  therefore reads
\begin{equation}
\langle \hat{O} \rangle = - 2\pi i |\tilde C_0|^2 \int \! \ d\Omega_5  \oint d\zeta   \,
\frac{1}{\zeta} \ 
\langle \Psi | \hat{O}   \, \hat{W}(\zeta,\phi) |\Psi(\zeta) \rangle  .
\end{equation}
The explicit form of the operator $\hat{W}$ is given in  Appendix \ref{app:2}. 

We can expand $W^{(0)}(\zeta, \phi)$ as a Laurent series in $\zeta$, with coefficients $d_n(\phi)$.
The normalization constants $\tilde C_n$ are related to integrated coefficients 
$\tilde d_n\equiv\int d\Omega_5 d_n(\phi) $
as in the pure number-projection case the $C_n$
are to the Laurent coefficients $d_n$ of $S^{(0)}(\zeta)$.

The forms of the expectation values of $\hat H_0$, $\hat H_{\rm int}$ and $\hat N$
are as given in the previous section, in Eqs~(\ref{eq:h02}) and (\ref{eq:hint2}), but setting $n=0$, and with
the replacement of  $d_n$ with $\tilde d_n$ and of the integrals ${\cal I}$ and ${\cal J}$ with 
the new integrals $I$ and $J$ which are given in  Appendix \ref{app:2}.  Once again
a one-parameter variation with respect to the gap $\Delta$ is performed.

\section{Results}
\label{sec:res}

\begin{figure}
\centerline{\includegraphics[width=8cm]{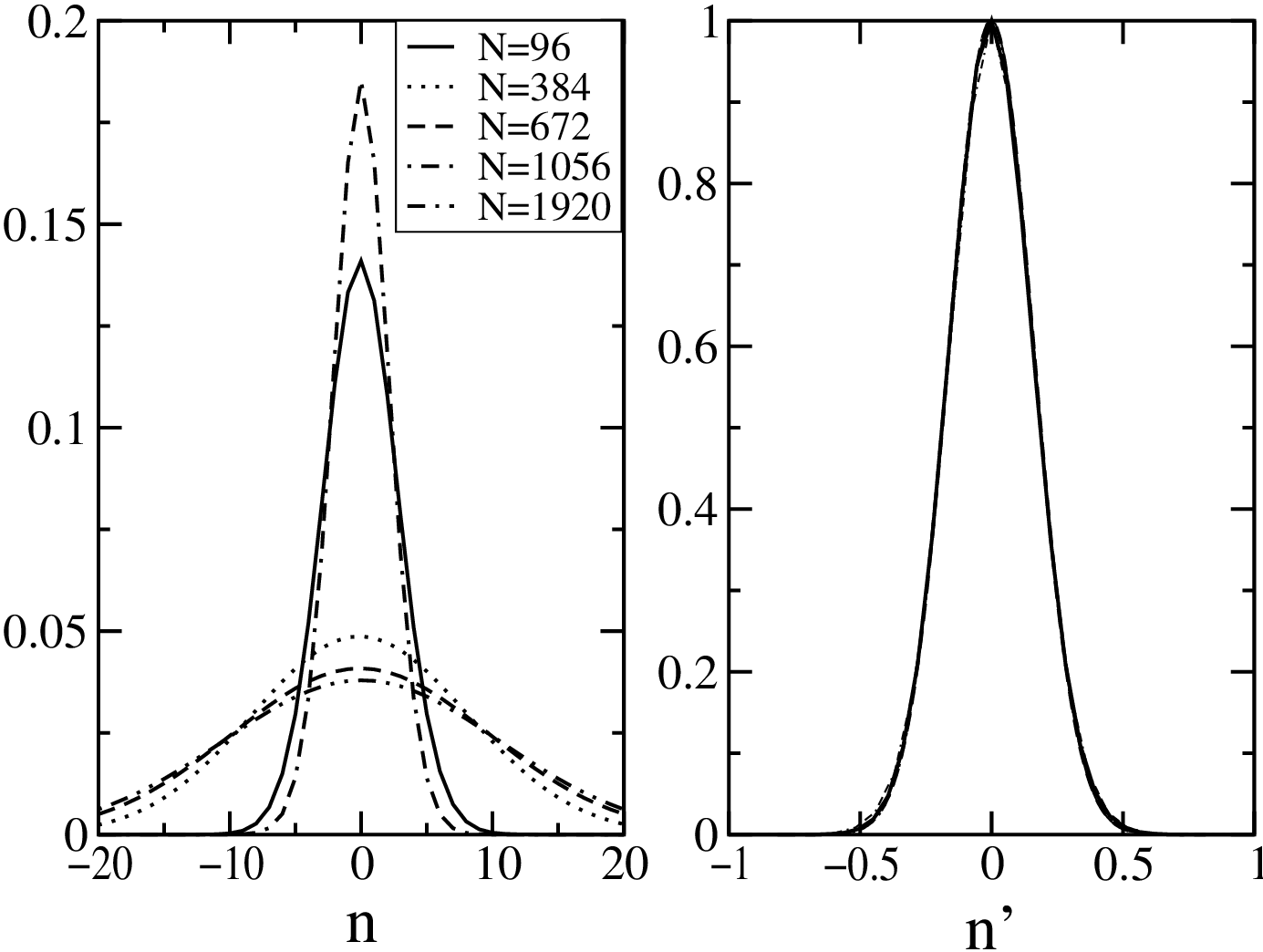}}
\caption{Left plot: distribution of the Laurent coefficients of $S^{(0)}(\zeta)$ for a box of size 
$L = 6 \text{ fm}$ and for different densities corresponding to the intersections between the two curves in 
the previous figure. The number of particles corresponding to each curve is  indicated. 
Right plot: distribution of the rescaled Laurent coefficients $d'_n = d_n/d_0$, as functions of the 
rescaled variable $n' = (d_0/\sqrt{2 \pi} )\ n$.}
\label{fig:paperfig7}
\end{figure}

\begin{figure*}[ptb]
\includegraphics[width=7cm]{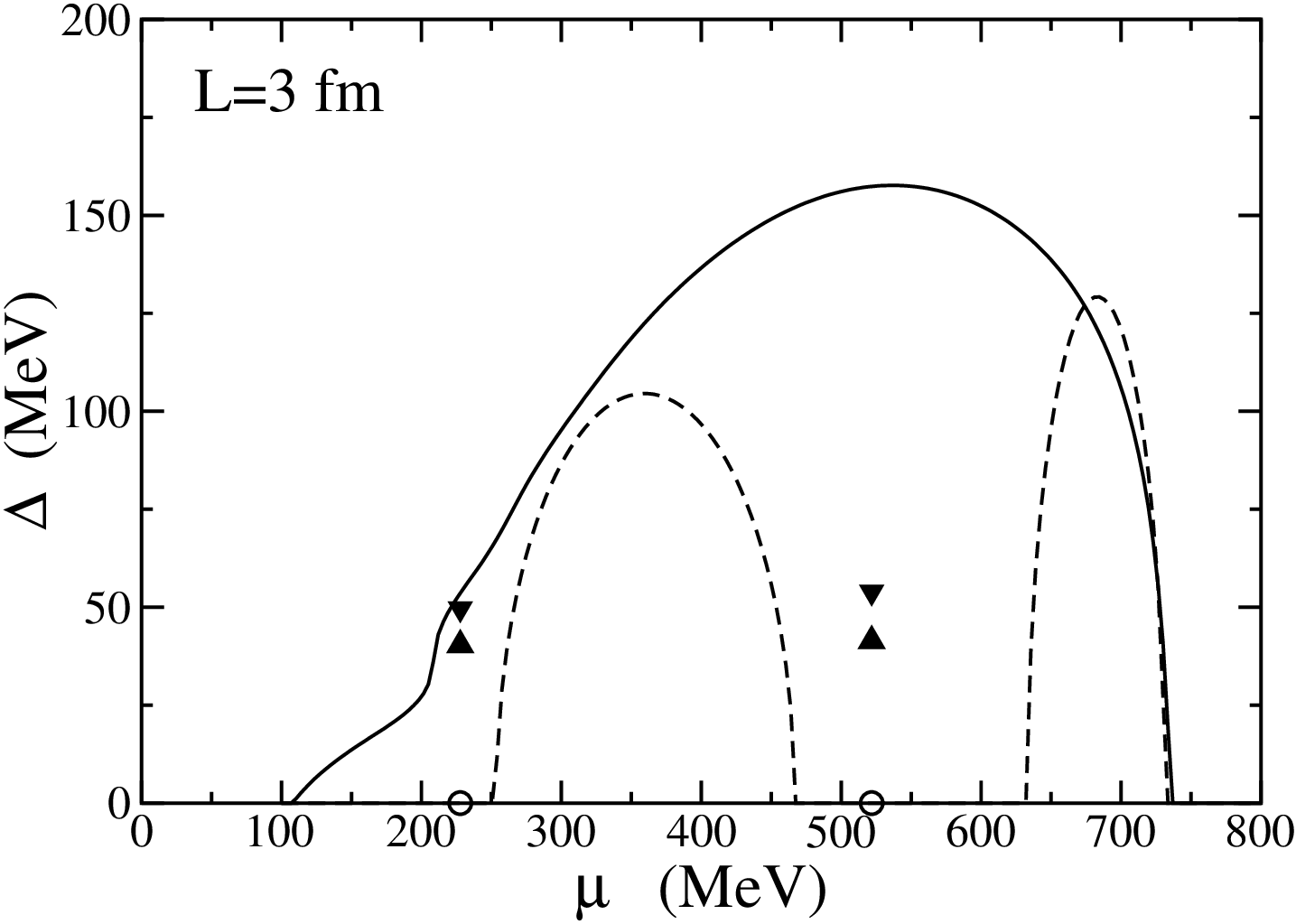} \hspace{1cm} \includegraphics[width=7cm]{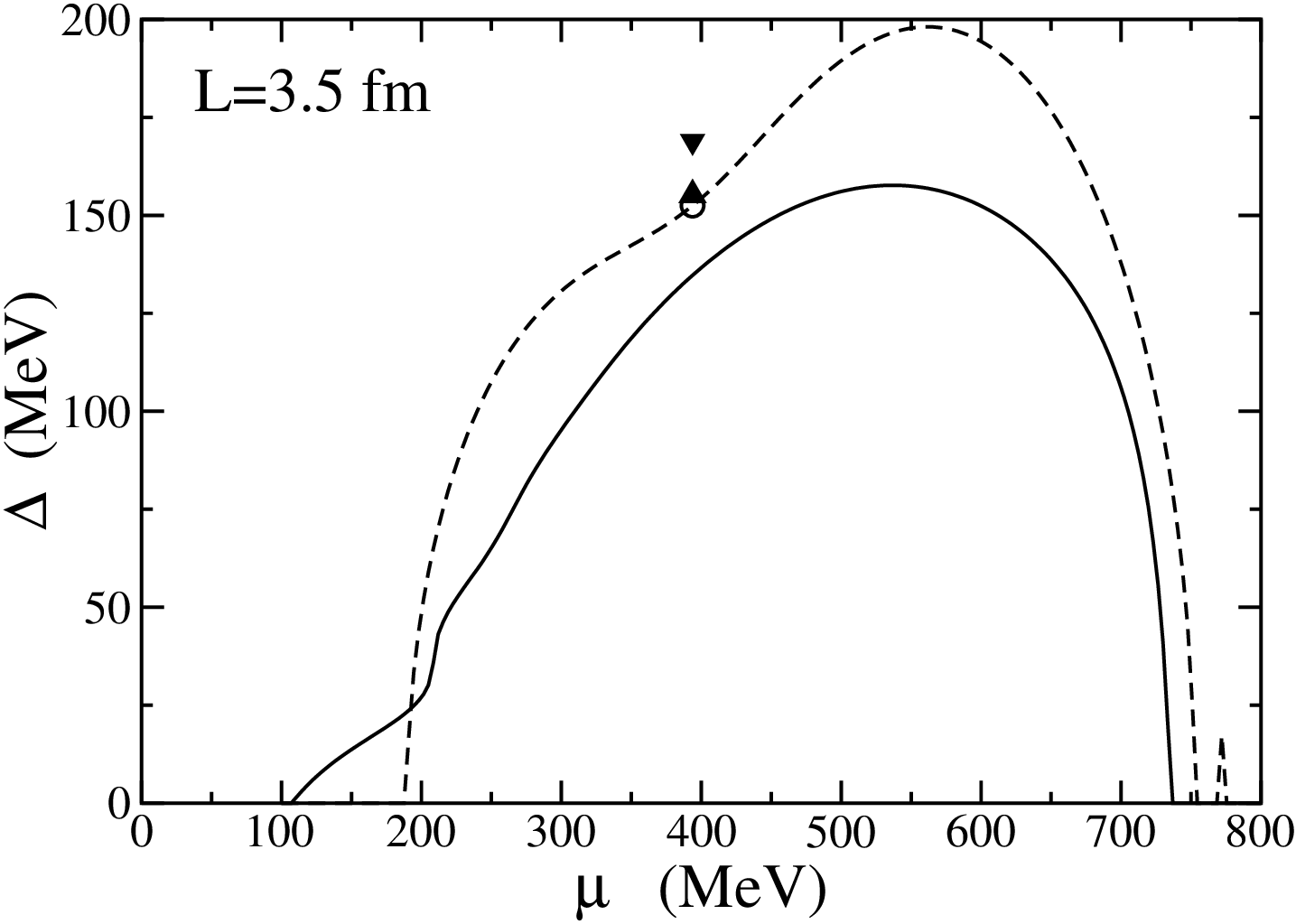}\\
\includegraphics[width=7cm]{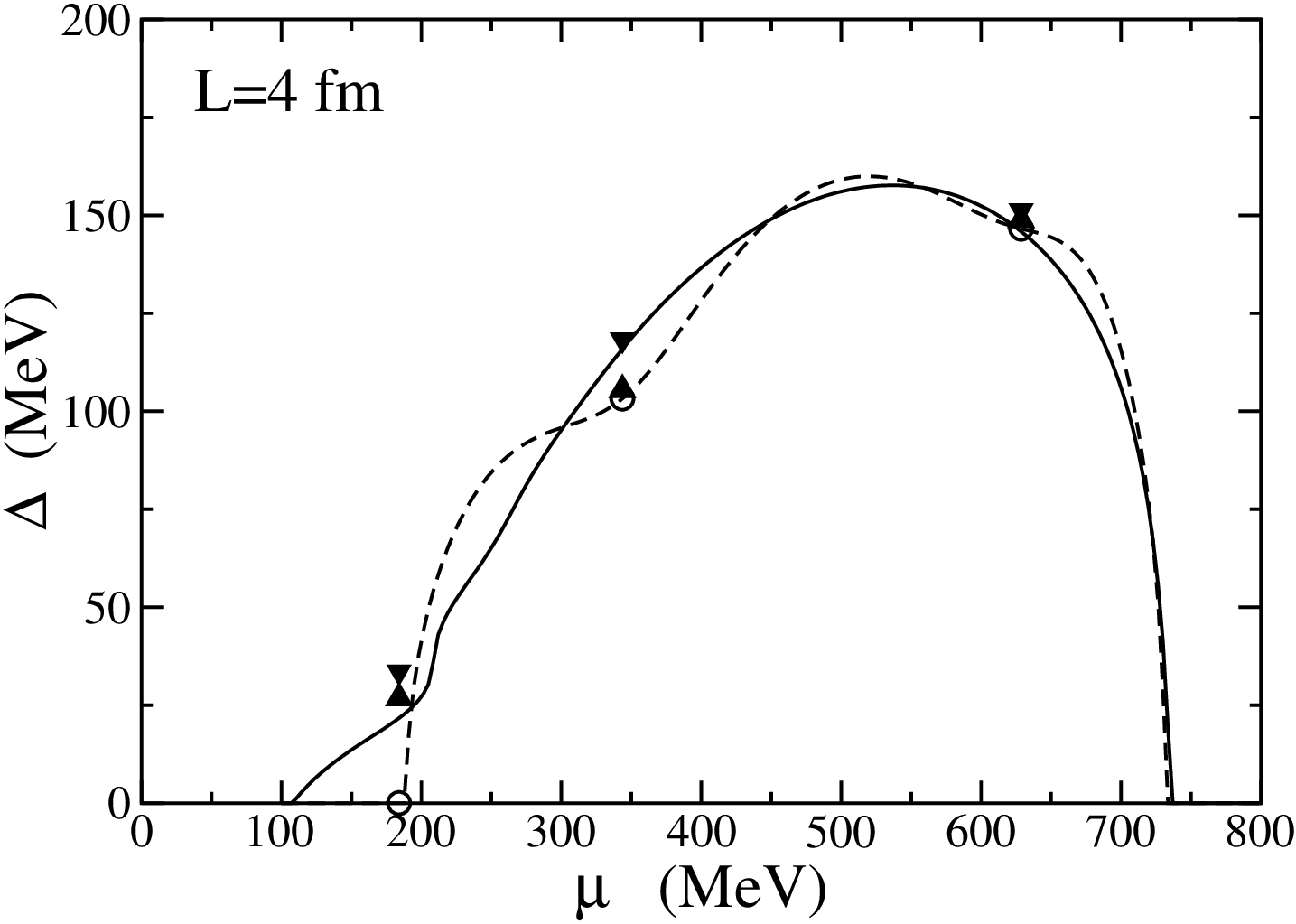} \hspace{1cm} \includegraphics[width=7cm]{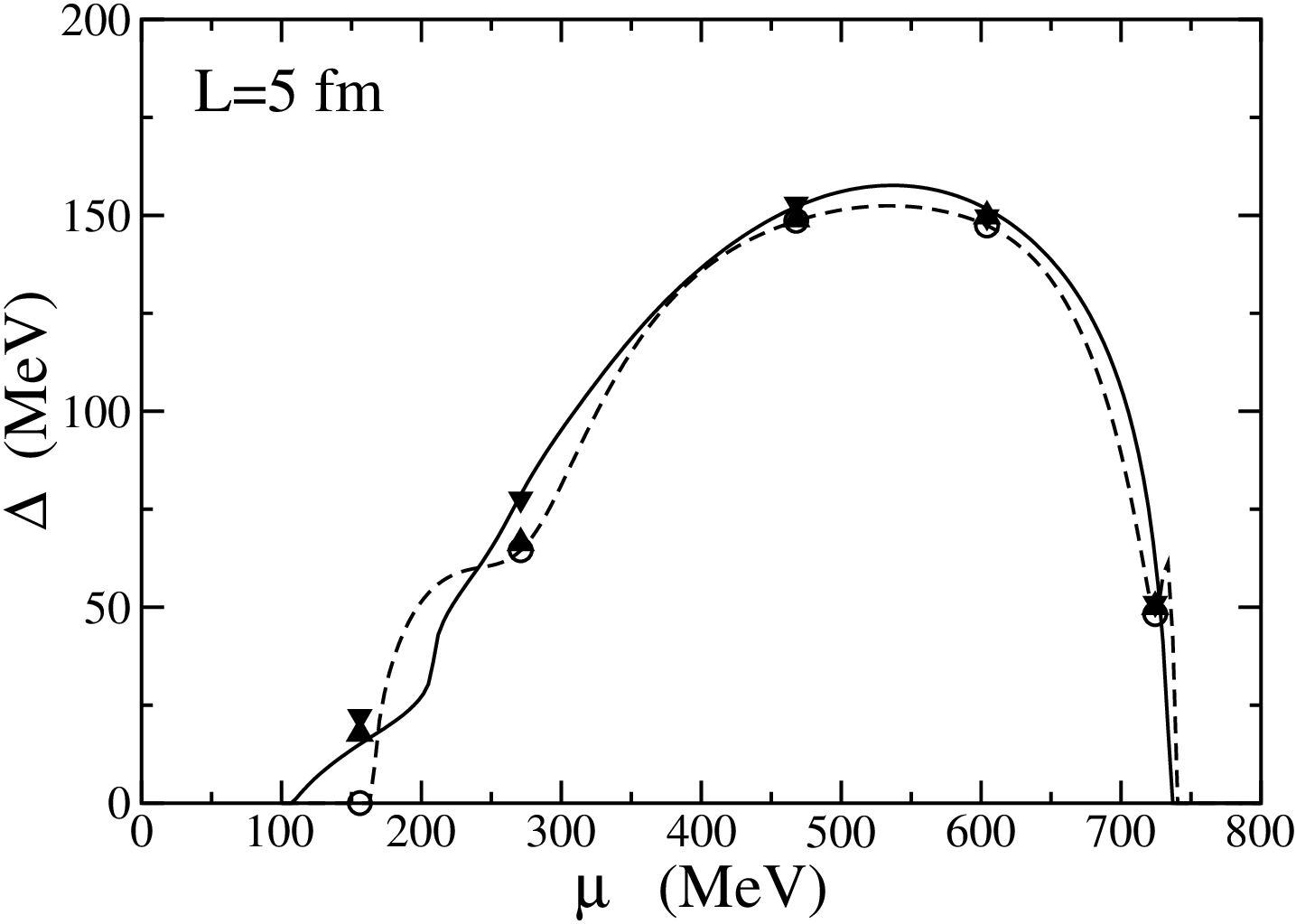}\\
\includegraphics[width=7cm]{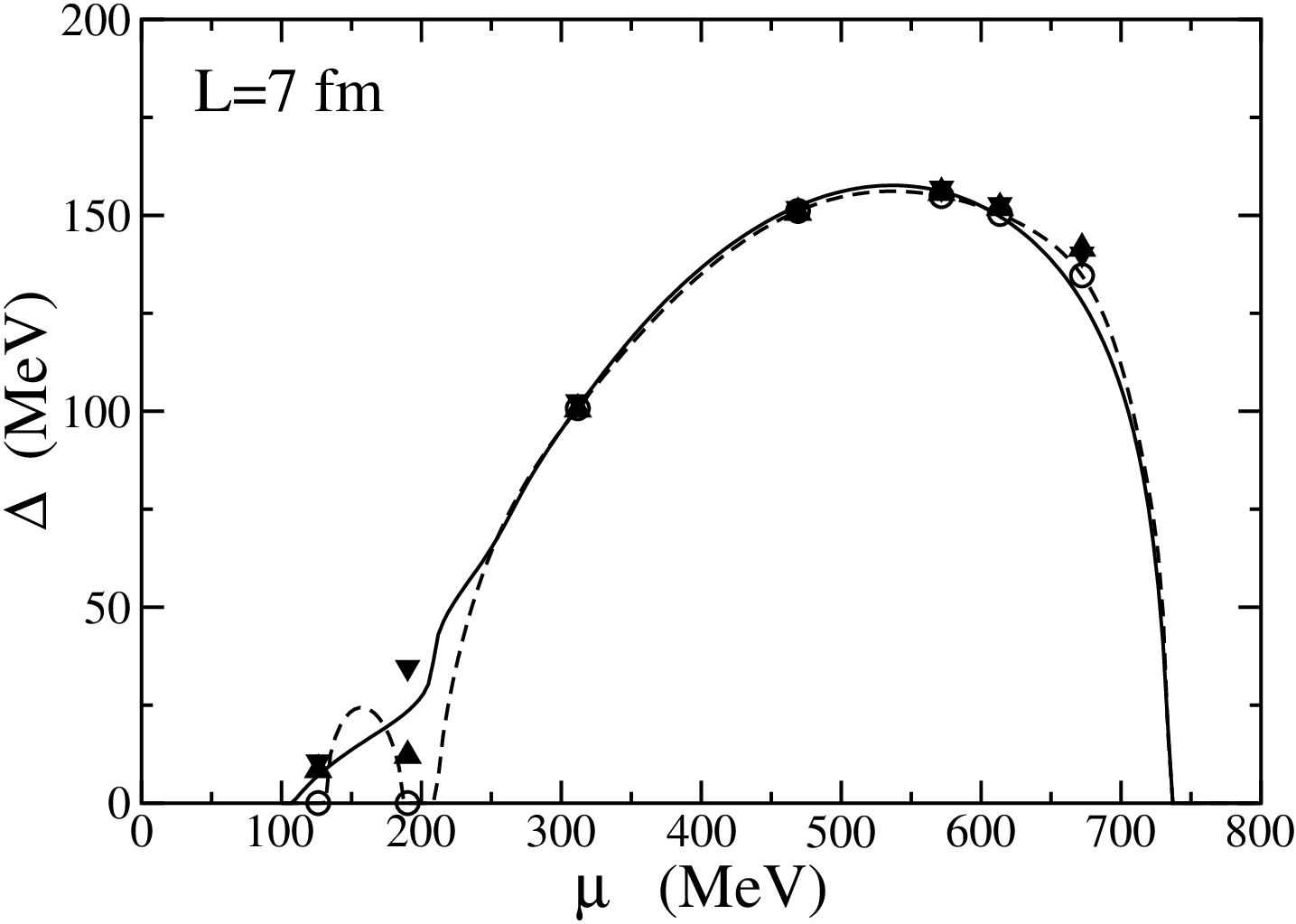} \hspace{1cm} \includegraphics[width=7cm]{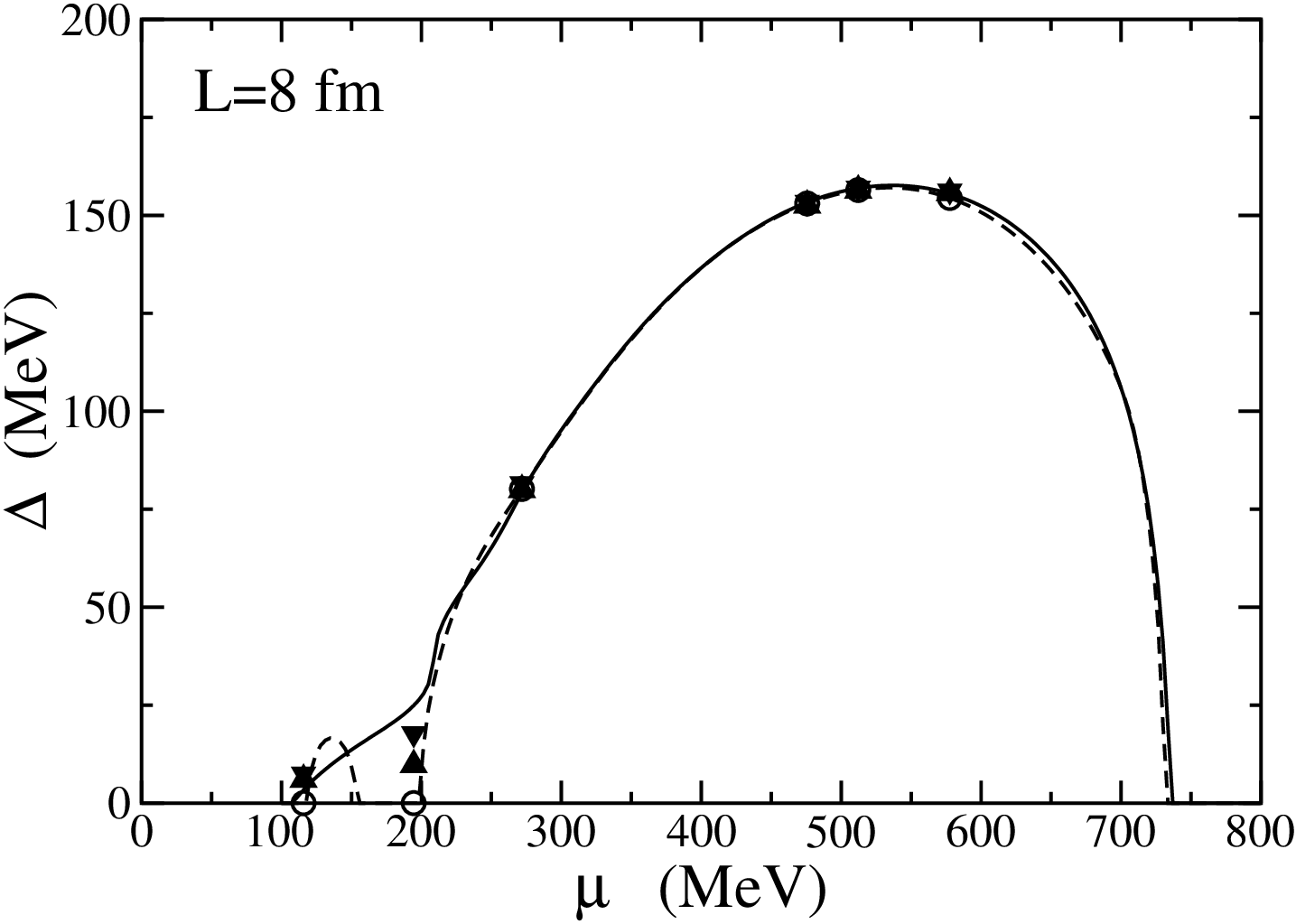}\\
\caption{Gap for the finite box at a fixed size of $L = 3,3.5,4,5,7,8 \ \text{fm}$: 
unprojected (circles) and dashed line, colour projected (cross), number projected (triangle up) and number-colour
projected (triangle down). The solid line is the gap for the infinite volume unprojected BCS state.}
\label{fig:gap_L3}
\end{figure*}

In this section we present our final results for the projected states.

First however we look at the effect of finite size on the unprojected BCS state. As the size 
of the box grows the fractional fluctuations in the baryon number will drop as $1/\sqrt{N}$,
and we expect that the value of the gap 
and of the energy should converge toward their values in the infinite volume. 
This behaviour is evident in Fig.~\ref{fig:fsize1}, which shows the superconducting gap
and total energy per particle. Convergence is obtained  quickly and continuously.

Now we turn to the effects of number and colour projection.
As we have explained in Section \ref{sec:colour_proj}, the projection over colour is carried 
out on BCS states having the same total baryon number as the underlying Fermi gas they are built on.
This restricts us to certain values of $\mu$, as we saw from 
Fig.~\ref{fig:paperfig6}. In Eq.~(\ref{eq:Thouless1}) we introduced the function $S^{(0)}(\zeta)$
whose Laurent coefficients $d_n$
are related to the amount of the projected state with $n$ pairs contained in the unprojected 
state (see Eq.~(\ref{eq:comp})).

Fig.~\ref{fig:paperfig7} shows the distribution of the Laurent coefficients for
the function $S^{(0)}(\zeta)$ at these points.Because we are working at our chosen values of $\mu$, 
the coefficients peak at
$n=0$, and we see that the unprojected BCS state has a substantial overlap with the projected state with 
zero pairs. The distributions are easily approximated with Gaussian 
functions.

In Fig.~\ref{fig:gap_L3} we show the superconducting gap as a function of chemical potential for a range 
of box sizes. The dashed line is the finite-volume unprojected gap, and for comparison 
the continuous line shows the result for the infinite-volume case.  We remind the reader that because
we perform only a limited variation after projection, as explained above, the projection can only be 
done for certain discrete values of $\mu$ for which the average net number of pairs in the unprojected
BCS state is zero. At these points we show  results for the  unprojected BCS 
(circles, which must coincide with the dashed line), and for the  
number-projected (triangle up) and  combined number-colour projected BCS (triangle down). 

The most striking finite-volume effect is the disappearance of the gap for values of $\mu$ which are
far from shells.  These regions are confined to very low  $\mu$ for boxes above 3~fm 
in length.  In these regions the effects of projection on the gap can be striking, as can be seen even
at 7~fm.  However in all other regions, where the unprojected gap does not disappear, projection has very
limited effect.  In general the results for colour and number projection lie above those for number projection
alone; the latter show almost no difference from the unprojected case unless the unprojected gap vanishes.

The effect of finite size on the energy of the condensate has already been shown in Fig.~\ref{fig:fsize1}.
The effect of projection is to lower the energy by less than 1~MeV even for the smallest boxes considered.
In the approach to the infinite-volume limit, shell effects for the unpaired quarks vanish more slowly
than finite-size or projection effects on the condensate.

\section{Conclusions}
\label{sec:concl}

In this work we have studied the effects of finite size over the colour superconducting state,
by considering quark matter at finite density and zero temperature enclosed in a cubic box.
We have considered the 2SC superconducting state of ref.~\cite{ARW98} and we have calculated the 
effects of the projection over definite baryon number and over a colour singlet state. 
The most significant finite size effect is the vanishing of the gap for values of the chemical potential
between widely spaced shells.  For these cases, projection has a significant effect, restoring a sizeable
gap.  Otherwise the effects of projection are slight, though in all cases each projection increases the gap
to some degree. 

This behaviour points to the interesting possibility that the interplay between the chiral  
and the superconducting phase can also change in a finite system.  This hypothesis can be verified 
by extending the present calculation, allowing a competition between the two phases.
This topic is currently under investigation \cite{ABDMW01}.

There are other aspects which require future investigation as well: it would be interesting to extend
our results to more general form of pairing interactions, rather than the simple ``instanton-motivated''
interaction of Eq.(\ref{eq:hint1}). For technical reasons, 
the number projection which was carried
out exactly in the present work, would require more effort in a general case, where the interaction 
contributions will contain not only terms which mix the helicity, such as $\overline{\psi}_L \ \psi_R$, 
but also terms which connect fermions with the same helicity, such as $\overline{\psi}_L \ \psi_L$.
It would be more appropriate in such cases to apply approximate number projection techniques, which have 
been developed in the contest of nuclear theory \cite{RS80}.

Another aspect of interest, which we plan to investigate in the future, is the effect of
the colour projection on different superconducting states, and in particular on the colour-flavour locked phase 
(CFL), which was originally studied in \cite{ARW99}. Because in this state the colour and flavour degrees of 
freedom are ``locked'', the colour projection will affect both colour and flavour, and could possibly 
yield larger effects than the one observed in the case of the 2SC.

We also remark that in the present analysis we have not performed a full variation after projection.
Instead we have followed a simpler approach, assuming for the solution the same functional form as 
in the unprojected case and obtaining the numerical solution by performing a minimization of 
the total energy with respect  to the gap  $\Delta$.

\begin{acknowledgments}
This work was supported by the UK EPSRC.
The authors would like to thank Dr.~M. Alford for stimulating discussions and suggestions. 
P.A. would like to thank Prof.~J.D. Walecka for useful discussions on the  topic.
\end{acknowledgments}

\begin{widetext}
\appendix

\section{Number projection: some useful formulas}
\label{app:1}

\subsection{Quasi-particle operators}
\label{app1:sub1}

We give here the explicit expressions for the quasi-particle operators annihilating 
the state   $|\psi(\zeta) \rangle$ of Eq.~(\ref{eq:wf2}). 
After some algebra we find that these can be written as follows:
\begin{eqnarray}
\hat{\alpha}_{L 1 \alpha}(\zeta,\kvec) &=& \cos\theta_A^L(k)  \hat{a}_{L 1 \alpha}(\kvec) -
\zeta  \epsilon_{\alpha \beta 3}  \sin\theta_A^L(k)  e^{i \xi_A^L(k)} \
\hat{a}_{L 2 \beta}^\dagger(-\kvec), \nonumber \\
\hat{\alpha}_{L 2 \beta}(\zeta,\kvec) &=& \cos\theta_A^L(k)  \hat{a}_{L 2 \beta}(-\kvec) +
\zeta  \epsilon_{\alpha \beta 3}  \sin\theta_A^L(k)  e^{i \xi_A^L(k)} \
\hat{a}_{L 1 \alpha}^\dagger(\kvec), \nonumber \\
\hat{\beta}_{R 1 \alpha}(\zeta,\kvec) &=& \cos\theta_B^L(k)  \hat{b}_{R 1 \alpha}(\kvec) -
\zeta^*  \epsilon_{\alpha \beta 3}  \sin\theta_B^R(k)  e^{i \xi_B^R(k)} \
\hat{b}_{R 2 \beta}^\dagger(-\kvec), \nonumber  \\
\hat{\beta}_{R 2 \beta}(\zeta,\kvec) &=& \cos\theta_B^R(k)  \hat{b}_{R 2 \beta}(-\kvec) +
\zeta^*  \epsilon_{\alpha \beta 3}  \sin\theta_B^R(k)  e^{i \xi_B^R(k)} \
\hat{b}_{R 1 \alpha}^\dagger(\kvec), \nonumber \\
\hat{\gamma}_{R 1 \alpha}(\zeta,\kvec) &=& \cos\theta_C^L(k)  \hat{c}_{R 1 \alpha}(\kvec) -
\zeta^*  \epsilon_{\alpha \beta 3}  \sin\theta_C^R(k)  e^{i \xi_C^R(k)} \
\hat{c}_{R 2 \beta}^\dagger(-\kvec), \nonumber  \\
\hat{\gamma}_{R 2 \beta}(\zeta,\kvec) &=& \cos\theta_C^R(k)  \hat{c}_{R 2 \beta}(-\kvec) +
\zeta^*  \epsilon_{\alpha \beta 3}  \sin\theta_C^R(k)  e^{i \xi_C^R(k)} \
\hat{c}_{R 1 \alpha}^\dagger(\kvec)  .
\end{eqnarray}

Notice that these operators fulfill canonical anticommutation relations. 
The inverse equations, which allow us to express the original operators in terms
of the quasi-particle ones, are also useful and read:
\begin{eqnarray}
\hat{a}_{L 1 \alpha}(\kvec) &=&  \cos\theta_A^L(k)  \hat{\alpha}_{L 1 \alpha}(\zeta,\kvec) +
\zeta  \epsilon_{\alpha \beta 3}  \sin\theta_A^L(k)  e^{i \xi_A^L(k)} \
\hat{\alpha}_{L 2 \beta}^\dagger(\zeta,\kvec), \nonumber  \\
\hat{a}_{L 2 \beta}(-\kvec) &=&  \cos\theta_A^L(k)  \hat{\alpha}_{L 2 \beta}(\zeta,\kvec) 
- \zeta  \epsilon_{\alpha \beta 3}  \sin\theta_A^L(k)  e^{i \xi_A^L(k)} \
\hat{\alpha}_{L 1 \alpha}^\dagger(\zeta,\kvec), \nonumber \\
\hat{b}_{R 1 \alpha}(\kvec) &=&  \cos\theta_B^R(k)  \hat{\beta}_{R 1 \alpha}(\zeta,\kvec) +
\zeta^*  \epsilon_{\alpha \beta 3}  \sin\theta_B^R(k)  e^{i \xi_B^R(k)} \
\hat{\beta}_{R 2 \beta}^\dagger(\zeta,\kvec), \nonumber  \\
\hat{b}_{R 2 \beta}(-\kvec) &=&  \cos\theta_B^R(k)  \hat{\beta}_{R 2 \beta}(\zeta,\kvec) 
-\zeta^*  \epsilon_{\alpha \beta 3}  \sin\theta_B^R(k)  e^{i \xi_B^R(k)} \
\hat{\beta}_{R 1 \alpha}^\dagger(\zeta,\kvec), \nonumber \\
\hat{c}_{R 1 \alpha}(\kvec) &=&  \cos\theta_C^R(k)  \hat{\gamma}_{R 1 \alpha}(\zeta,\kvec) +
\zeta^*  \epsilon_{\alpha \beta 3}  \sin\theta_C^R(k)  e^{i \xi_C^R(k)} \
\hat{\gamma}_{R 2 \beta}^\dagger(\zeta,\kvec), \nonumber  \\
\hat{c}_{R 2 \beta}(-\kvec) &=&  \cos\theta_C^R(k)  \hat{\gamma}_{R 2 \beta}(\zeta,\kvec) 
-\zeta^*  \epsilon_{\alpha \beta 3}  \sin\theta_C^R(k)  e^{i \xi_C^R(k)} \
\hat{\gamma}_{R 1 \alpha}^\dagger(\zeta,\kvec) . 
\end{eqnarray}

\subsection{Thouless operator}
\label{app1:sub2}

We now consider the operator defined in Eq.~(\ref{eq:Thouless1}). 
For the part corresponding to $\hat{G}_L^\dagger(\zeta,\kvec)$ 
we look for solutions satisfying
\begin{eqnarray}
\left[ w_1^A(\zeta) + \epsilon_{\alpha \beta 3}  w_2^A(\zeta)  
\hat{a}^\dagger_{L 1 \alpha}(\kvec)  \hat{a}^\dagger_{L 2 \beta}(-\kvec) \right]
\hat{G}_{L A \alpha \beta}^\dagger(1,\kvec) &=& 
\hat{G}_{L A \alpha \beta}^\dagger(\zeta,\kvec), \nonumber \\
\left[ w_1^B(\zeta) + \epsilon_{\alpha \beta 3}  w_2^B(\zeta)  
\hat{b}^\dagger_{R 1 \alpha}(\kvec)  \hat{b}^\dagger_{R 2 \beta}(-\kvec) \right]
\hat{G}_{L B \alpha \beta}^\dagger(1,\kvec) &=& 
\hat{G}_{L B \alpha \beta}^\dagger(\zeta,\kvec), \nonumber \\
\left[ w_1^C(\zeta) + \epsilon_{\alpha \beta 3}  w_2^C(\zeta)  
\hat{c}^\dagger_{R 1 \alpha}(\kvec)  \hat{c}^\dagger_{R 2 \beta}(-\kvec) \right]
\hat{G}_{L C \alpha \beta}^\dagger(1,\kvec) &=& 
\hat{G}_{L C \alpha \beta}^\dagger(\zeta,\kvec) .
\end{eqnarray}
These have the forms
\begin{eqnarray}
w_1^A(\zeta) = 1 ,\qquad    &&    w_2^A(\zeta) = 
(\zeta-1)  \tan \theta_A^L(k)  e^{i \xi_A^L(k)},\nonumber \\
w_1^B(\zeta) = 1   ,\qquad &&     w_2^B(\zeta) = 
\left(\zeta^*-1\right)  
\tan\theta_B^R(k)  e^{i \xi_B^R(k)},\nonumber \\
w_1^C(\zeta) = 1   ,\qquad &&    w_2^C(\zeta) = 
\left(\zeta^*-1\right)  
\tan\theta_C^R(k)  e^{i \xi_C^R(k)} .
\end{eqnarray}

By simple substitutions we can also obtain the similar solutions for the part 
corresponding to  $\hat{G}_R^\dagger(\zeta,\kvec)$. 

Finally the Thouless operator is:
\begin{eqnarray}
\hat{S}(\zeta)=\hat{S}_{L}(\zeta) \hat{S}_{R}(\zeta) 
 &=& \hat{S}_{L A}(\zeta)   \hat{S}_{R B}(\zeta) \
 \hat{S}_{R C}(\zeta) \hat{S}_{R A}(\zeta)  
\hat{S}_{L B}(\zeta)  \hat{S}_{L C}(\zeta) \ ,
\label{eq:Thouless2}
\end{eqnarray}
where 
\begin{eqnarray}
\hat{S}_{L A}(\zeta)  &=& \prod_{\absum\kvec} 
\left[ 1 + (\zeta-1)  \epsilon_{\alpha \beta 3}   \tan \theta_A^L(k)  e^{i \xi_A^L(k)} \
\hat{a}^\dagger_{L 1 \alpha}(\kvec)  \hat{a}^\dagger_{L 2 \beta}(-\kvec) \right], \nonumber \\
 \hat{S}_{R B}(\zeta) &=& 
\prod_{\absum\kvec} \left[ 1 + \left(\zeta^*-1\right)  \epsilon_{\alpha \beta 3}  
\tan\theta_B^R(k)  e^{i \xi_B^R(k)}  \hat{b}^\dagger_{R 1 \alpha}(\kvec)  
\hat{b}^\dagger_{R 2 \beta}(-\kvec) \right], \nonumber \\
 \hat{S}_{R C}(\zeta) &=& \prod_{\absum\kvec} \left[ 1 + \left(\zeta^*-1\right)  
\epsilon_{\alpha \beta 3} \
\tan\theta_C^R(k)  e^{i \xi_C^R(k)}
\hat{c}^\dagger_{R 1 \alpha}(\kvec)  \hat{c}^\dagger_{R 2 \beta}(-\kvec) \right], \nonumber \\
\hat{S}_{R A}(\zeta) &=& \prod_{\absum\kvec} \left[ 1 + (\zeta-1)  \epsilon_{\alpha \beta 3}  
\tan \theta_A^R(k)  e^{i \xi_A^R(k)} \
\hat{a}^\dagger_{R 1 \alpha}(\kvec)  \hat{a}^\dagger_{R 2 \beta}(-\kvec) \right], \nonumber \\
\hat{S}_{L B}(\zeta) 
&=& \prod_{\absum\kvec} \left[ 1 + \left(\zeta^*-1\right)  \epsilon_{\alpha \beta 3}  
\tan\theta_B^L(k)  e^{i \xi_B^L(k)}  \hat{b}^\dagger_{L 1 \alpha}(\kvec)  
\hat{b}^\dagger_{L 2 \beta}(-\kvec) \right], \nonumber \\
\hat{S}_{L C}(\zeta) 
&=& \prod_{\absum\kvec} \left[ 1 + \left(\zeta^*-1\right)  \epsilon_{\alpha \beta 3} \
\tan\theta_C^L(k)  e^{i \xi_C^L(k)}
\hat{c}^\dagger_{L 1 \alpha}(\kvec)  \hat{c}^\dagger_{L 2 \beta}(-\kvec) \right]  .
\end{eqnarray}

To apply Wick's theorem it is preferable to express the equations above directly in terms of
the quasi-particle operators:
\begin{eqnarray}
\hat{S}_{L A}(\zeta)  &~& \hskip-.5cm | \Psi \rangle = 
 \prod_{\absum\kvec} \left[ \left( \cos^2\!\theta_A^L(k) + \zeta  
\sin^2 \!\theta_A^L(k) \right) \right. \nonumber \\
&+& \left.  (\zeta-1)  \epsilon_{\alpha \beta 3}   \sin\theta_A^L(k) 
\cos\theta_A^L(k) e^{i \xi_A^L(k)}  \hat{\alpha}_{L 1 \alpha}^\dagger(1,\kvec) 
\hat{\alpha}_{L 2 \beta}^\dagger(1,\kvec) \right] | \Psi \rangle ,
\nonumber \\
\hat{S}_{R A}(\zeta) &~& \hskip-.5cm  | \Psi \rangle 
= \prod_{\absum\kvec} \left[ \left( \cos^2\!\theta_A^R(k) + \zeta  \sin^2 \!\theta_A^R(k) \right) 
\right. \nonumber \\
&+& \left.  (\zeta-1)  \epsilon_{\alpha \beta 3}   \sin\theta_A^R(k) \cos\theta_A^R(k) 
e^{i \xi_A^R(k)}  \hat{\alpha}_{R 1 \alpha}^\dagger(1,\kvec) \hat{\alpha}_{R 2 \beta}^\dagger(1,\kvec) 
\right] | \Psi \rangle, \nonumber \\
\hat{S}_{R B}(\zeta) &~& \hskip-.5cm  | \Psi \rangle 
= \prod_{\absum\kvec} \left[ \left( \cos^2\!\theta_B^R(k) + \zeta^*   \sin^2 \!\theta_B^R(k) \right) 
\right. \nonumber \\
&+& \left.  \left(\zeta^*-1\right)  \epsilon_{\alpha \beta 3}   
\sin\theta_B^R(k) \cos\theta_B^R(k) e^{i \xi_B^R(k)}  \hat{\beta}_{R 1 \alpha}^\dagger(1,\kvec) 
\hat{\beta}_{R 2 \beta}^\dagger(1,\kvec)  \right] | \Psi \rangle, \nonumber \\
\hat{S}_{L B}(\zeta) &~& \hskip-.5cm  | \Psi \rangle 
= \prod_{\absum\kvec} \left[ \left( \cos^2\!\theta_B^L(k) + \zeta^*  \sin^2 \!\theta_B^L(k) \right) 
\right. \nonumber \\
&+& \left.  \left(\zeta^*-1\right)  \epsilon_{\alpha \beta 3}   
\sin\theta_B^L(k) \cos\theta_B^L(k) e^{i \xi_B^L(k)}  \hat{\beta}_{L 1 \alpha}^\dagger(1,\kvec) 
\hat{\beta}_{L 2 \beta}^\dagger(1,\kvec)  \right] | \Psi \rangle, \nonumber \\
\hat{S}_{R C}(\zeta) &~& \hskip-.5cm  | \Psi \rangle 
= \prod_{\absum\kvec} \left[ \left( \cos^2\!\theta_C^R(k) + \zeta^*  \sin^2 \!\theta_C^R(k) \right) 
\right. \nonumber \\
&+& \left.  \left(\zeta^*-1\right)  \epsilon_{\alpha \beta 3}   
\sin\theta_C^R(k) \cos\theta_C^R(k) e^{i \xi_C^R(k)}  \hat{\gamma}_{R 1 \alpha}^\dagger(1,\kvec) 
\hat{\gamma}_{R 2 \beta}^\dagger(1,\kvec)  \right] | \Psi \rangle, \nonumber \\
\hat{S}_{L C}(\zeta) &~& \hskip-.5cm  | \Psi \rangle 
= \prod_{\absum\kvec} \left[ \left( \cos^2\!\theta_C^L(k) + \zeta^*  \sin^2 \!\theta_C^L(k) \right) 
\right. \nonumber \\
&+& \left.  \left(\zeta^*-1\right)  \epsilon_{\alpha \beta 3}   
\sin\theta_C^L(k) \cos\theta_C^L(k) e^{i \xi_C^L(k)}  \hat{\gamma}_{L 1 \alpha}^\dagger(1,\kvec) 
\hat{\gamma}_{L 2 \beta}^\dagger(1,\kvec)  \right] | \Psi \rangle .
\end{eqnarray}

\subsection{Integrals}
\label{app1:sub3}

We have seen in Section \ref{sec:numb_proj} that the matrix elements of operators in 
number projected states can be expressed in terms of a limited number of integrals, namely
\begin{eqnarray}
{\cal J}_{a,n}(\theta_1,\theta_2) &\equiv&  \frac{1}{2 \pi i}  \oint  \frac{d\zeta}{\zeta^{n+1}}  
 S^{(0)}(\zeta)  \frac{\zeta}{( \cos^2\!\theta_1 + \zeta \sin^2 \!\theta_1 )( \cos^2\!\theta_2 + \zeta \sin^2 \!\theta_2 )}, 
\nonumber \\
{\cal J}_{b,n}(\theta_1,\theta_2) &\equiv&   \frac{1}{2 \pi i}  \oint  \frac{d\zeta}{\zeta^{n+1}}  
 S^{(0)}(\zeta)  \frac{\zeta^*}{(\cos^2\!\theta_1 + \zeta^* \sin^2 \!\theta_1 )  
(\cos^2\!\theta_2 + \zeta^* \sin^2 \!\theta_2 )},  
\nonumber \\
{\cal J}_{c,n}(\theta_1,\theta_2) &\equiv&   \frac{1}{2 \pi i}  \oint  \frac{d\zeta}{\zeta^{n+1}}  
 S^{(0)}(\zeta)  \frac{1}{( \cos^2\!\theta_1 + \zeta^* \sin^2 \!\theta_1 )( \cos^2\!\theta_2 + \zeta \sin^2 \!\theta_2 )}. 
\label{eq:app1J} 
\end{eqnarray}

The integrals for the single particle operators are obtained from those above:
\begin{eqnarray}
{\cal I}_{(a,b,c),n}(\theta) &\equiv& {\cal J}_{(a,b,c),n}(\theta,0)
\end{eqnarray}

\section{colour projection}
\label{app:2}

\subsection{Thouless operator}
\label{app2:sub1}

We can now give the explicit expression for the Thouless operator $\hat{W}$, which can be
derived in a similar fashion to $\hat{S}$. Following the same approach as before 
we first look for an operator, $\hat{T}$, that is at most bilinear in the quark operators
and that it can be applied to $\hat{G}(\kvec,\zeta)$ to give the colour rotated 
operator $\widetilde{{G}}(\kvec,\zeta,\phi)$.

For example, in order to obtain the operators in Eq.~(\ref{eq:rot2}) we 
need
\begin{eqnarray}
\hat{t}_{L A 1 2}(\kvec,\phi) &\equiv& 1 + w_1\,  \hat{a}^\dagger_{L 1 1}(\kvec)  \hat{a}_{L 1 1}(\kvec)+
w_2 \, \hat{a}^\dagger_{L 1 3}(\kvec)  \hat{a}_{L 1 1}(\kvec) \nonumber \\
\hat{t}_{L A 2 1}(\kvec,\phi) &\equiv& 1 + w_3 \, \hat{a}^\dagger_{L 2 1}(-\kvec)  \hat{a}_{L 2 1}(-\kvec) +
w_4 \, \hat{a}^\dagger_{ L 2 3}(-\kvec)  \hat{a}_{L 2 1}(-\kvec)  \ .
\end{eqnarray}

The conditions
\begin{eqnarray}
\hat{t}_{L A 1 2}(\kvec,\phi)\,  \hat{G}_{L A 1 2}(\kvec,\zeta) | k_F \rangle 
&=&  \widetilde{{G}}_{L A 1 2}(\kvec,\zeta,\phi)| k_F \rangle  \nonumber\\
\hat{t}_{L A 2 1}(\kvec,\phi) \, \hat{G}_{L A 2 1}(\kvec,\zeta) | k_F \rangle 
&=& \widetilde{{G}}_{L A 2 1}(\kvec,\zeta,\phi) | k_F \rangle . 
\end{eqnarray}
yield the solutions 
\begin{eqnarray}
w_1 = w_3 = \cos\frac{\phi}{2} - 1    ,\qquad    w_2 = w_4 = - \sin\frac{\phi}{2}   .
\end{eqnarray}

Finally one can write the total operator $\hat{T}$ in the factorised form:
\begin{eqnarray}
\hat{T}(\phi) &\equiv&  \prod_{\absum\kvec} \hat{t}_{L A \alpha \beta}(\kvec,\phi) \, 
\hat{t}_{L B \alpha \beta}(\kvec,\phi)\,  \hat{t}_{L C \alpha \beta}(\kvec,\phi) \
\hat{t}_{R A \alpha \beta}(\kvec,\phi)\,  \hat{t}_{R B \alpha \beta}(\kvec,\phi)\,  
\hat{t}_{R C \alpha \beta}(\kvec,\phi) \ . 
\end{eqnarray}

As a result, the Thouless operator for the combined colour and number projection will be obtained
by the application of $\hat{T}$ and $\hat{S}$:
\begin{eqnarray}
\hat{W}(\zeta,\phi)  &\equiv& \hat{T}(\phi)  \hat{S}(\zeta)  \nonumber \\
&=& \prod_{\absum\kvec}  
\hat{w}_{L A \alpha \beta}(\kvec)  \hat{w}_{L B \alpha \beta}(\kvec)  
\hat{w}_{L C \alpha \beta}(\kvec)  \hat{w}_{R A \alpha \beta}(\kvec)  
\hat{w}_{R B \alpha \beta}(\kvec)  \hat{w}_{R C \alpha \beta}(\kvec) \ ,
\end{eqnarray}

We will  simply state the results for the component
corresponding to left handed particles:
\begin{eqnarray}
\hat{W}_{L A}(\phi,\zeta) &=& \hat{W}_{L A}^{(0)}(\phi,\zeta) + \hat{W}_{L A}^{(1)}(\phi,\zeta) 
+ \dots  
\end{eqnarray}
where
\begin{eqnarray}
\hat{W}_{L A}^{(0)}(\phi,\zeta) &=& \prod_{\absum\kvec} \
\left(\cos^2\!\theta_A^L(k) + \zeta  \cos\frac{\phi}{2}  \sin^2 \!\theta_A^L(k) \right)  
\end{eqnarray}
and
\begin{eqnarray}
\hat{W}_{L A}^{(1)}(\phi,\zeta) &=& \hat{W}_{L A}^{(0)}(\phi,\zeta) \nonumber \\
&\times& \left\{\sum_{\gamma \delta \kvec} 
\frac{\epsilon_{\gamma \delta 3}  \left( \zeta  \cos\frac{\phi}{2} - 1 \right)  
\sin\theta_A^L(k)  \cos\theta_A^L(k)  e^{i \xi_A^L(k)}}{
\left(\cos^2\!\theta_A^L(k) + \zeta  \cos\frac{\phi}{2}  \sin^2 \!\theta_A^L(k) \right)}
 \hat{\alpha}_{L 1 \gamma}^\dagger(1,\kvec)  \hat{\alpha}_{L 2 \delta}^\dagger(1, \kvec)
\right. \nonumber \\
&-&\left. \sum_{\kvec}\frac{\zeta \sin\theta_A^L(k)   
\sin\frac{\phi}{2}  e^{i \xi_A^L(k)}}{\left(\cos^2\!\theta_A^L(k) + \zeta  
\cos\frac{\phi}{2}  \sin^2 \!\theta_A^L(k) \right)} 
\left(\hat{a}^\dagger_{L 1 3}(\kvec)\hat{\alpha}_{L 2 2}^\dagger(1, \kvec)+
\hat{a}^\dagger_{L 2 3}(\kvec)\hat{\alpha}_{L 1 2}^\dagger(1, \kvec)\right)\right\}.
\end{eqnarray}
The last term pairs the third colour with one of the
other two. This term does not to contribute to the expectation value of the
Hamiltonian.

\subsection{Integrals}
\label{app2:sub2}

In analogy with  Sec. \ref{app1:sub3} we now define the following objects:
\begin{eqnarray}
J_{a,n}(\theta_1,\theta_2) &\equiv& \frac{1}{2 \pi i}  \oint \frac{d\zeta}{\zeta^{n+1}}  \int \! d\Omega_5 
 W^{(0)}(\zeta,\phi)  \frac{\zeta \cos\frac{\phi}{2}}{(\cos^2\!\theta_1+\sin^2 \!\theta_1  \zeta \cos\frac{\phi}{2})  
(\cos^2\!\theta_2+\sin^2 \!\theta_2  \zeta \cos\frac{\phi}{2})},\nonumber\\
J_{b,n}(\theta_1,\theta_2) &\equiv& \frac{1}{2 \pi i}  \oint \frac{d\zeta}{\zeta^{n+1}}  \int \! d\Omega_5
 W^{(0)}(\zeta,\phi)  \frac{\zeta^* \cos\frac{\phi}{2}}{(\cos^2\!\theta_1+\sin^2 \!\theta_1  \zeta^* \cos\frac{\phi}{2})  
(\cos^2\!\theta_2+\sin^2 \!\theta_2  \zeta^* \cos\frac{\phi}{2}) },\nonumber \\
J_{c,n}(\theta_1,\theta_2) &\equiv& \frac{1}{2 \pi i}  \oint \frac{d\zeta}{\zeta^{n+1}}  \int \! d\Omega_5
 W^{(0)}(\zeta,\phi)  \frac{1}{(\cos^2\!\theta_1+\sin^2 \!\theta_1  \zeta^* \cos\frac{\phi}{2})   
(\cos^2\!\theta_2+\sin^2 \!\theta_2  \zeta \cos\frac{\phi}{2})},
\label{app2:J}
\end{eqnarray}
where $d\Omega_5$ is given by Eq.~(\ref{eq:red}). These reduce to the integrals ${\cal J}$ of Eq.~(\ref{eq:app1J})
if $\phi$ is set to zero and the colour-group integral is suppressed.

The integrals for single particle operators will be obtained from the ones above as a particular case:
\begin{eqnarray}
I_{(a,b,c) , n}(\theta) &\equiv& J_{(a,b,c) , n}(\theta,0)  .
\end{eqnarray}

In order to calculate these integrals, we write the Laurent series for $W^{(0)}(\zeta,\phi)$ as
\begin{eqnarray}
W^{(0)}(\zeta,\phi) = \sum_n  d_n(\phi)  \zeta^n  \ .
\end{eqnarray}

Numerically it is found that the Laurent coefficients  as a function of colour angle
$\phi$  can be fit by  Gaussians with a single parameter $\alpha$:
\begin{eqnarray}
d_n(\phi) &\approx& d_n(0)  e^{-\frac{\alpha}{2}  \sin^2 \!\frac{\phi}{2} } \ .
\label{eq:alpha}
\end{eqnarray}

This approximation is particularly useful in the numerical calculation, since it allows us to 
calculate the integrals analytically. With this simplification, 
the integrals in Eq.~(\ref{app2:J}) can be written in a closed form as:
\begin{eqnarray}
J_{a,n}(\theta_1,\theta_2) &=& \sum_m  d_m(0)  \left[ I_1  \Lambda_{n,m}(\theta_1,\theta_2) \right. \nonumber \\
&+& \left. 
\frac{I_2}{2}  
\left( \sin^2 \!\theta_1  \sin^2 \!\theta_2  \Delta_{n,m+2}(\theta_1,\theta_2)
- \cos^2\!\theta_1  \cos^2\!\theta_2  \Delta_{n,m}(\theta_1,\theta_2) \right)\right], \nonumber \\
J_{b,n}(\theta_1,\theta_2) &=&  \sum_m  d_m(0)  \left[ I_1  
\Lambda_{n,m}(\theta^{\rm c}_1,\theta^{\rm c}_2) \right. \nonumber \\
&+& \left.  \frac{I_2}{2}   \left( \sin^2 \!\theta_1  \sin^2 \!\theta_2  \Delta_{n,m}(\theta^{\rm c}_1,\theta^{\rm c}_2)
- \cos^2\!\theta_1  \cos^2\!\theta_2  \Delta_{n,m+2}(\theta^{\rm c}_1,\theta^{\rm c}_2)\right) \right], \nonumber \\
J_{c,n}(\theta_1,\theta_2) &=& \sum_m  d_m(0)  \left[ I_1  
\Lambda_{n,m}(\theta^{\rm c}_1,\theta_2) \right. \nonumber \\
&+& \left. \frac{I_2}{2}  
\left( \cos^2\!\theta_1  \sin^2 \!\theta_2  \Delta_{n,m+2}(\theta^{\rm c}_1,\theta_2) + 
\sin^2 \!\theta_1  \cos^2\!\theta_2  \Delta_{n,m}(\theta^{\rm c}_1,\theta_2) \right. \right. \nonumber \\
&+& \left. \left. 2  \sin^2 \!\theta_1  \sin^2 \!\theta_2 \
\Delta_{n,m+1}(\theta^{\rm c}_1,\theta_2) \right) \right]  , 
\end{eqnarray}
where $\theta^{\rm c}= \frac{1}{2}\pi-\theta$ and we have defined
\begin{eqnarray}
\Lambda_{n,m}(\theta_1,\theta_2) &\equiv& \frac{1}{2 \pi i}  \oint \frac{d\zeta}{\zeta^{n-m}} \
\frac{1}{(\cos^2\!\theta_1+\sin^2 \!\theta_1  \zeta )  (\cos^2\!\theta_2+\sin^2 \!\theta_2  \zeta)}, \nonumber\\
\Delta_{n,m}(\theta_1,\theta_2) &\equiv& \frac{1}{2 \pi i}  \oint \frac{d\zeta}{\zeta^{n-m}} \
\frac{1}{(\cos^2\!\theta_1+\sin^2 \!\theta_1  \zeta )^2  (\cos^2\!\theta_2+\sin^2 \!\theta_2  \zeta)^2} \ .
\label{eq:lam}
\end{eqnarray}
and
\begin{eqnarray}
I_1 &\equiv& \int \! d\Omega_5 \ e^{- \frac{\alpha}{2} \sin^2 \!\frac{\phi}{2}} = \frac{2 - (\alpha+2) \
e^{-\frac{\alpha}{2}}}{\alpha^2}\nonumber \\
I_2 &\equiv& \int \! d\Omega_5 \, \sin^2 \!\frac{\phi}{2} \ e^{- \frac{\alpha}{2}\sin^2 \!\frac{\phi}{2}}=
\frac{8 - (\alpha^2+4 \alpha + 8) \  e^{-\frac{\alpha}{2}}}{\alpha^3}.
\end{eqnarray}
The parameter $\alpha$ is defined in Eq.~(\ref{eq:alpha}).
The integrals of Eq.~(\ref{eq:lam}) can be also evaluated analytically.

\bigskip

\end{widetext}


\end{document}